\documentclass[12pt]{iopart}

\usepackage{epsfig}
\usepackage{graphicx}
\usepackage{bm}
\usepackage{iopams}
\usepackage[sort]{cite}

\expandafter\let\csname equation*\endcsname\relax

\expandafter\let\csname endequation*\endcsname\relax

\usepackage{amsmath}
\usepackage{lscape}
\usepackage[usenames,dvipsnames]{color} 
\usepackage[normalem]{ulem}

\usepackage{dcolumn}
\usepackage{bm}

\begin{document}

\title[Tuning Pairwise Potential Can Control the Fragility of Glass-Forming Liquids]{Tuning Pairwise Potential Can Control the Fragility of Glass-Forming Liquids: From Tetrahedral Network to Isotropic Soft Sphere Models}

\author{Misaki Ozawa$^{1, 2}$, Kang Kim$^{3, 4}$, and Kunimasa Miyazaki$^1$}
\address{$^1$Department of Physics, Nagoya University, Nagoya 464-8602, Japan }
\address{$^2$ Laboratoire Charles Coulomb, UMR 5221 CNRS-Universit\'e de Montpellier, Montpellier, France} 
\address{$^3$Department of Physics, Niigata University, Niigata
950-2181, Japan}
\address{$^4$Division of Chemical Engineering, Graduate School of Engineering Science, Osaka University, Toyonaka, Osaka 560-8531, Japan}
\ead{miyazaki@r.phys.nagoya-u.ac.jp}
\date{\today}

\begin{abstract}
We perform molecular dynamics simulations for a $\mathrm{SiO_2}$ glass former
 model proposed by Coslovich and Pastore (CP) over a wide range of density.
The density variation can be mapped onto the change of the
 potential depth between Si and O interactions of the CP model. 
By reducing the potential depth (or increasing the density), 
the anisotropic tetrahedral network structure observed in the original CP model
transforms into the isotropic structure with the purely repulsive
 soft-sphere potential.
Correspondingly, the temperature dependence of the 
 relaxation time exhibits the crossover from the
 Arrhenius to the super-Arrhenius behavior.
Being able to control the fragility over a wide range by tuning the potential of a single model system helps us to bridge the gap between the network and isotropic glass formers
and to obtain the insight into the underlying mechanism of the fragility.
We study the relationship between the fragility and dynamical
properties such as the magnitude of the Stokes-Einstein
 violation and the stretch exponent in the density correlation  function.
We also demonstrate that the peak of the specific heat systematically shifts as the density increases,
hinting that the fragility is correlated with the hidden thermodynamic anomalies of the system.
  
\end{abstract}

\maketitle

\section{INTRODUCTION}
\label{sec:INTRODUCTION}

The current understanding of the mechanism behind the dramatic slowing down of supercooled liquids near the
glass transition temperatures still remains far from complete. 
Although underlying mechanism of the glass transition is believed to be universal, many dynamical
properties of the glass formers are diverse and system-dependent. 
Among them, the concept of the fragility, or the temperature dependence of the relaxation time and 
transport coefficients such as the viscosity and diffusion constant, is one of the most important but
the least understood problems. 
For some glass formers, which are called the {\it strong} liquids
according to the Angell's classification, 
the relaxation times obey the Arrhenius law,
while others, called the {\it fragile} liquids, show a strong departure from the Arrhenius
law in their temperature dependence of the relaxation times~\cite{angell1995formation,greer2014fragility}.  
Representative strong glass formers are 
$\rm SiO_2$ and $\rm GeO_2$, whose local molecular configurations are characterized by
anisotropic network structures. On the other hand, the fragile liquids such as $o$-terphenyl and
toluene tend to have more isotropic and compact local structures.
The concept of the fragility has been playing a key role in the study of the glass transition.
Many experimental studies have reported that there exist 
correlations between the fragility and various thermodynamic, structural, mechanical, and dynamical
properties of the glass formers~\cite{greer2014fragility}.
For example, it has been observed that the entropy and specific heat~\cite{angell1995formation,martinez2001thermodynamic}, the 
elastic constants~\cite{novikov2004poisson,novikov2005correlation}, and
the spatially heterogeneous dynamics~\cite{bohmer1993nonexponential,niss2007correlation} sensitively depend on the fragility of the systems. 
Theoretical understanding of the fragility, on the other hand, is left at the phenomenological level.
Virtually all theoretical scenarios of the glass transition proposed so far, including 
the Adam-Gibbs theory\cite{adam1965temperature}, the random first order transition theory~\cite{xia2000fragilities}, energy landscape picture~\cite{debenedetti2001supercooled},
geometric frustration scenarios~\cite{tarjus2000viscous,
coslovich2007understanding1, tanaka2012bond, royall2015strong}, elastic models~\cite{dyre2006colloquium,krausser2015interatomic}, 
soft modes~\cite{yan2013glass}, and  the dynamic
facilitation scenario~\cite{garrahan2003coarse}, have their own explanations of the fragility but
the first-principles and microscopic theory to quantitatively describe the fragility is still
lacking~\cite{debenedetti2001supercooled,cavagna2009supercooled,berthier2011theoretical,binder2011glassy,royall2015role}.

Difficulty  to gain an unified picture of the fragility out
of the experimental observations is obviously
due to the diversity and complexity of real
molecular or polymeric glass formers.
The computer simulation is an ideal tool to avoid this difficulty because one can gain 
all microscopic information of dynamics for relatively simple model systems.
Popular glass former models in the simulation studies, such as the soft-sphere (SS)~\cite{bernu1987soft} and the Lennard-Jones (LJ) potential
liquids~\cite{coslovich2007understanding1,wahnstrom1991molecular,kob1995testing1}, are regarded as typical fragile glass formers.
On the other hand, strong glass former models with anisotropic networks, where the particles are connected by
covalent bonds, such as $\rm SiO_2$ and $\rm GeO_2$, have also been studied extensively by
simulations~\cite{woodcock1976molecular,tsuneyuki1988first,van1990force,coslovich2009dynamics}.   

There have been several attempts to understand the origin of the fragility
systematically by simulations.  
In many studies, the isotropic potential systems such as the SS and LJ potentials have been employed.
The fragility has been controlled using various methods, for example,
by varying the density or pressure of the system~\cite{sastry2001relationship,de2004scaling,sengupta2013density,berthier2009compressing,wang2012non},
by changing the polydispersity~\cite{kawasaki2007correlation,kawasaki2010structural,abraham2008energy}
and the size ratio of the multi-component mixtures~\cite{coslovich2007understanding1,kurita2010glass},  
by tuning the interaction potential~\cite{de2004scaling,bordat2004does,sengupta2011dependence},
or by truncating the attractive part of the LJ potential~\cite{berthier2009nonperturbative, coslovich2011locally}. 
The fragility was also found to be sensitive to many-body interactions which influence the local
geometrical frustration~\cite{shintani2006frustration,molinero2006tuning}.
An idea to sort out thermodynamic and purely kinetic contributions to the
fragility has been discussed recently~\cite{parmar2015kinetic}. 
The external parameters such as the impurities~\cite{kim2011slow,chakrabarty2015dynamics}
and the curvature of the non-Euclidean space~\cite{sausset2008tuning} have been proposed as a new
method to control the fragility. 
The findings of these studies, however, are still not sufficient to unravel the entire picture of the fragility.
Aside from the obvious drawback that the time window which the current simulation can
cover is narrow compared with experiments, the one of the obstacles of computational studies is that 
the variation of the fragility which the simple glass former models can encompass is
small~\cite{sastry2001relationship,sengupta2013density,kawasaki2007correlation, kawasaki2010structural,coslovich2007understanding1,kurita2010glass, abraham2008energy,de2004scaling, bordat2004does, sengupta2011dependence,berthier2009nonperturbative,
coslovich2011locally}. 
Furthermore, the variation of the fragility is often masked by an apparent density-temperature scaling law, which makes it difficult to extract the generic mechanism controlling the fragility~\cite{de2004scaling,sengupta2013density}. There also remains the nagging question of how to decompose the kinetic and purely thermodynamic contributions of the fragility~\cite{parmar2015kinetic}. 
Even more important is to bridge the gap between the fragile isotropic systems and the
strong network glass formers, which has been studied almost in different arenas in the past.
Given these circumstances, it is beneficial to consider a simple model
system which can cover a wide range of fragility, spanning from the strongest 
network glass former down to the very fragile isotropic system, simply by tuning a system parameter. 

In this paper, we numerically study a simple model glass former originally introduced by
Coslovich and Pastore  (CP) as a model strong glass former mimicking $\rm
SiO_2$~\cite{coslovich2009dynamics}. 
The CP model is a binary mixture of spherical atoms interacting by the isotropic Lennard-Jones
potential with a very strong attraction between Si and O atoms, which allows them to 
form anisotropic tetrahedral network structures.
In this study, the fragility of the CP
model is examined over an extremely wide
range of density from the order of unity up
to virtually infinity.
Varying the density is equivalent with changing the potential depth between Si and O atoms of the CP model.
By reducing this potential depth (or increasing the density), 
the anisotropic tetrahedral network structure observed in the CP model
 is eventually transformed into the isotropic structure with the purely repulsive
 soft-sphere potential.
Thus, it enables one to control the fragility from strong to fragile systematically.

Here, we summarize the results of this paper.
By extensive simulations, dependence of the structure on the density is monitored using the radial distribution
functions, the static structure factors, and the coordination numbers.
By measuring the temperature dependence of the relaxation time obtained from the density-density time correlation
function, we observe that the fragility systematically changes from strong to fragile 
with the change of the structures.  
We carefully examine the density-temperature scaling for the temperature dependence of the
relaxation time~\cite{alba2002temperature, casalini2004thermodynamical,alba2006temperature, pedersen2008strong,schroder2009pressure3, gnan2009pressure4, schroder2011pressure5}.
It is confirmed that the relaxation time 
does not collapse onto a master curve.
This implies that the observed fragility variation is not superficial but
it is due to a qualitative change of the underlying mechanism
controlling the slow dynamics.
We investigate correlation between the fragility and several dynamical observables, {\it i.e.}, the magnitude of the
Stokes-Einstein (SE) violation and the exponent of the stretched exponential
relaxation of the density correlation function.
It has been argued that these observables are manifestations of spatially
heterogeneous dynamics, or {\it the dynamical heterogeneities}, and they are intimately correlated
to the fragility~\cite{ediger2000spatially}.
We confirm that more fragile systems show stronger SE
violation, whereas a clear-cut correlation between the stretch exponent and the fragility is not observed.
Finally, we also discuss the possibility that the fragility is related to the peak position of the
specific heat observed in its temperature dependence.

This paper is organized as follows.
In Section~\ref{sec:SIMULATION_METHODS}, we describe the details of the
simulation methods.
The numerical results are presented in
Section~\ref{sec:RESULTS}.
Finally, we discuss our results and conclude in Section~\ref{sec:SUMMARY_AND_DISCUSSION}.

\section{SIMULATION METHODS}
\label{sec:SIMULATION_METHODS}

In this section, we describe the simulation methods.

\subsection{The Coslovich-Pastore model}
Coslovich and Pastore (CP) have introduced a simple $\rm SiO_2$ (strong glass
former) model~\cite{coslovich2009dynamics}.
Its dynamical and structural properties have been intensively investigated for fixed
densities~\cite{berthier2012finite,kim2013multiple,kawasaki2014dynamics,staley2015reduced}.  
The advantage of this model is its simplicity compared to other established $\rm SiO_2$
models~\cite{woodcock1976molecular,tsuneyuki1988first,van1990force}. 
Although its potential is given by a combination of the soft-sphere (SS) and the Lennard-Jones (LJ) potentials, the atoms in the CP model form an
anisotropic tetrahedral network structure similar to those of the 
realistic $\rm SiO_2$ models~\cite{coslovich2009dynamics}. 
According to References~\cite{zaccarelli2007spherical,mayer2010spherical},
there are three conditions for a binary mixture with spherical potential to generate the tetrahedral network structure;
(1) The composition ratio is $N_1:N_2 = 1:2$, where $N_1$ and $N_2$ are
the number of particles for species 1 and 2, respectively.
(2) The potential is non-additive, {\it i.e}, 
the range of the interaction between the species 1 and 2 is not a simple sum of their diameters.
(3) The attractive interaction between different species is strong.
The CP model satisfies these conditions.

The CP model is a binary mixture in three dimensions whose composition
ratio is $N_1 : N_2 = 1 : 2$ and the mass ratio is $m_2/m_1=0.57$. 
The species 1 and 2 correspond to $\rm Si$ and $\rm O$ atoms of $\rm SiO_2$, respectively.
The interaction potentials between two particles are given by
\begin{equation}
v_{\alpha_i \beta_j}(r)=\epsilon_{\alpha_i \beta_j} \left\{
						     \left(\frac{\sigma_{\alpha_i
						      \beta_j}}{r}\right)^{12}
						     -
						     C(1-\delta_{\alpha_i
						     \beta_j}) \left(
								\frac{\sigma_{\alpha_i
								\beta_j}}{r}
							       \right)^{6}
						    \right\},
\label{eq:potential}
\end{equation}
where $C$ is a constant which was set to unity in the original CP model~\cite{coslovich2009dynamics}. 
We specify the particles by the Roman indices $i, j \in \{1, 2, \cdots, N\}$ and the
species by the Greek indices $\alpha_i, \beta_j \in \{1,2\}$. 
The parameters are set to
$\epsilon_{12}/\epsilon_{11}=24$, $\epsilon_{22}/\epsilon_{11}=1$, 
$\sigma_{22}/\sigma_{11}=0.85$, and $\sigma_{12}/\sigma_{11}=0.49$.
The last condition warrants that the potential is non-additive, {\it i.e.}, $\sigma_{12}\neq (\sigma_{11}+\sigma_{22})/2$.
$v_{\alpha_i \beta_j}(r)$ is truncated at $r=2.5\sigma_{\alpha \beta}$.
In order to ensure the continuity of the potential, we add a switching function, $S(r)$,  up to $r=3\sigma_{\alpha \beta}$.
The switching function connects two points, $R_1$ and $R_2$,
continuously and is defined as
\begin{equation}
S(r) = 
\left\{
\begin{aligned}
\quad &1  \qquad (0 \leq r < R_1),   \\
\quad &(r-R_2)^2 (2r+R_2 -3R_1)/(R_2 - R_1)^3   \quad \quad (R_1 \leq r < R_2), \\
\quad &0  \qquad (R_2 \leq r).
\end{aligned}
\right.
\end{equation}
We used $R_1=2.5\sigma_{\alpha \beta}$ and $R_2=3\sigma_{\alpha \beta}$.

\subsection{Tuning of the potential depth}

\begin{figure}
\begin{center}
\includegraphics[width=0.95\columnwidth]{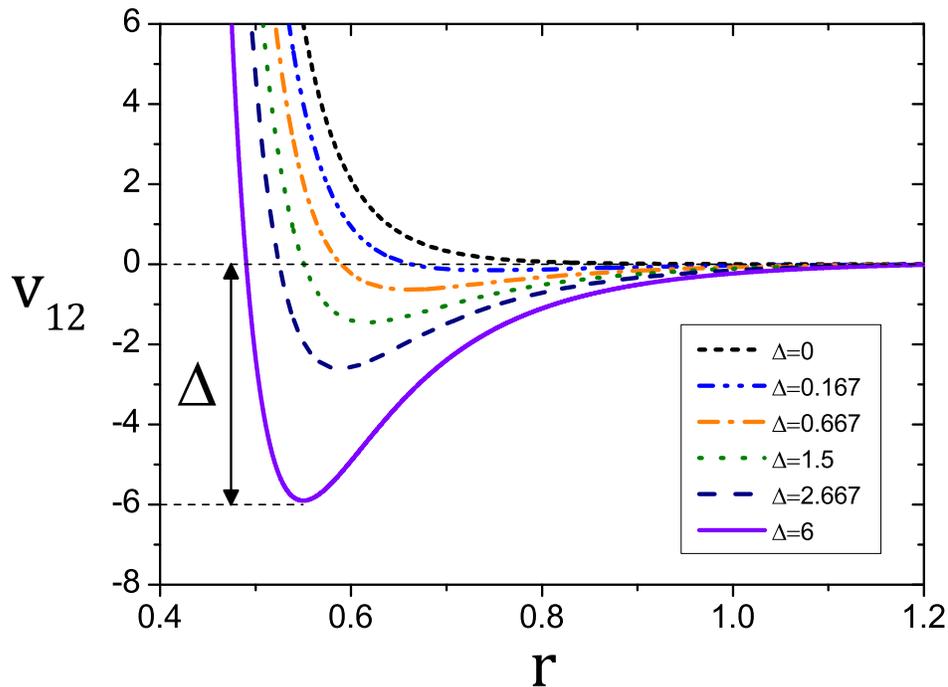}
\caption{
The interaction potential $v_{12}(r)$ between species 1 and 2 for 
various $\Delta$'s from $\Delta=0$ (top) to $\Delta=6$
 (bottom).
$\Delta=6$ corresponds to the original CP model in Reference~\cite{coslovich2009dynamics}.
}
\label{fig:potentials}
\end{center}
\end{figure}

In this study, we introduce a parameter $C$ in the second term (the
attraction part) of $v_{12}(r)$ in Equation~(\ref{eq:potential}).
$C$ is the strength of the attraction between different species.
The depth of the potential well $\Delta$ is written in the unit of $\epsilon_{11}$ as $\Delta=\epsilon_{12}C^2 /4\epsilon_{11}$.
Hereafter, we use $\Delta$ instead of $C$. 
In Figure~\ref{fig:potentials}, we show $v_{12}(r)$ for various $\Delta$'s.
$\Delta=6$ ($C=1$) corresponds to the original CP model and $\Delta=0$ ($C=0$) corresponds to a simple
soft-sphere (SS) potential.
Note that the potential for $\Delta=0$ is slightly different from the conventional SS potential
model which
has been extensively studied in the past~\cite{bernu1987soft}, since the range of the
interaction is still non-additive.
Our model can seamlessly connect the original CP model to the purely repulsive SS potential model.
As mentioned above, the atoms with large $\Delta$, \textit{i.e.}, the strong attractive
interaction, tend to form the tetrahedral network structure.
This structure is broken as $\Delta$ reduces and eventually 
the local structure becomes isotropic and compact in the limit of $\Delta = 0$.  
Mathematically, changing $\Delta$ at a constant density $\rho$ is equivalent to changing $\rho$ at a constant $\Delta$.
We explain the relation between $\Delta$ and $\rho$ using a simple
scaling argument, following the similar argument for the SS potential model~\cite{hiwatari1974molecular}.
The Hamiltonian of the system is written as
\begin{equation}
  \frac{H}{k_{\rm B} T}=\frac{1}{k_{\rm B} T}
\left\{\sum_{i=1}^{N}\frac{{\bm p}_i^2}{2m_{\alpha_i}} + \sum_{i<j} v_{\alpha_i \alpha_j}(r_{ij})\right\},
\end{equation}
where $k_{\rm B}$, ${\bm p}_i$, $m_{\alpha_i}$, and $v_{\alpha_i \alpha_j}(r_{ij})$ are the Boltzmann constant, momentum, mass, and the potential given by Equation (\ref{eq:potential}), respectively.
Introducing the unit of the length $l = \rho^{-1/3}$, this Hamiltonian can be rewritten as
\begin{equation}
 \begin{aligned}
\frac{H}{{k_{\rm B} T}}  &=& \Gamma^{4}\left[
		     \sum_{i=1}^{N} \frac{{{\bm p}_i^*}^2}{2m_{\alpha_i}^*}  +  \sum_{i<j} \epsilon_{\alpha_i
		     \alpha_j}^* \left\{ \left(\frac{\sigma_{\alpha_i
					  \alpha_j}^*}{r_{ij}^*}\right)^{12}
-C \left(\rho \sigma_{11}^3 \right)^{-2}
	    (1-\delta_{\alpha_i \alpha_j}) \left( \frac{\sigma_{\alpha_i\alpha_j}^*}{r_{ij}^*}
					  \right)^{6} \right\} \right], 
 \end{aligned}
\label{eq:scaling}
\end{equation}
where $m_{\alpha_i}^*$, ${\bm p}_{\alpha_i}^*$, $\sigma_{\alpha_i\alpha_j}^*$, and 
$\epsilon_{\alpha_i \alpha_j}^* $ are reduced parameters scaled by 
$m_{1}$, $m_1l/\tau_0$, $\sigma_{11}$, and $\epsilon_{11}$, respectively.
The time unit is defined by $\tau_0=\{ m_1 l^2/\epsilon_{11}(l/\sigma_{11})^{12}\}^{1/2}$. 
$\Gamma \equiv \rho (\epsilon_{11}/k_{\rm B}T)^{1/4}\sigma_{11}^3$ is a dimensionless parameter commonly 
used
for the SS potential~\cite{hiwatari1974molecular}.
Note that there exist two dimensionless parameters which control the
system: $\Gamma$ and $C \left(\rho \sigma_{11}^3 \right)^{-2}$. 
This means that, for a fixed $\Gamma$ (or the temperature), 
changing the density with a fixed $C$ is equivalent with changing $C$ for a fixed density. 
Therefore, if one chooses the purely Lennard-Jones system ($C=1$) with a density $\rho_0$ as a reference system, 
one can infer thermodynamic and dynamical properties at arbitrary densities by tuning $C$ for the
fixed $\rho_0$.  
The mapping from $C$ to $\rho$ is given by $(\rho_0/\rho)^2 = C$, or equivalently in terms of
$\Delta$ by

\begin{equation}
\left( \frac{\rho_0}{\rho} \right)^2 = \sqrt{\frac{\Delta}{6}}. 
\label{eq:mapping2}
\end{equation}
From this mapping, one can cover the whole range of density up to $\rho=\infty$ by tuning $C$
(or $\Delta$) from a finite value down to 0.

\subsection{Details of molecular dynamics simulations}

We perform the molecular dynamics (MD) simulations
for various $\Delta$'s.
We use $\Delta=6$, $2.667$, $1.5$, $1.042$, $0.667$, $0.375$, $0.167$,
$0.042$ and $0$ at a constant density $\rho_0 = 1.655$.
Correspondence of various $\Delta$'s for a fixed density $\rho_0$ to various densities for a fixed $\Delta=6$ 
is summarized in Table~\ref{tab:parameters}.
By tuning $\Delta$, we explore a wide range of density from $1.655$ up to infinity of the original CP model.
The number of the particles is $N = N_1 + N_2 = 3000$ with $N_1 : N_2 = 1 : 2$.  
The temperature range which we perform the simulation are listed in
Table~\ref{tab:parameters}.
Hereafter, we use $\sigma_{11}$, $\epsilon_{11}/k_B$, and
$\sqrt{m_1\sigma_{11}^2/\epsilon_{11}}$, as the units of length, time, and temperature, respectively.
The micro-canonical MD method with the periodic boundary condition is
used to produce the trajectories.
A time step $\Delta t = 0.0005$ is chosen throughout this study.
For the calculations of the dynamic and thermodynamic quantities, we
average over four independent simulation runs.

\section{RESULTS}
\label{sec:RESULTS}

\subsection{Structural properties}

\begin{figure}
\begin{center}
\includegraphics[width=0.48\columnwidth]{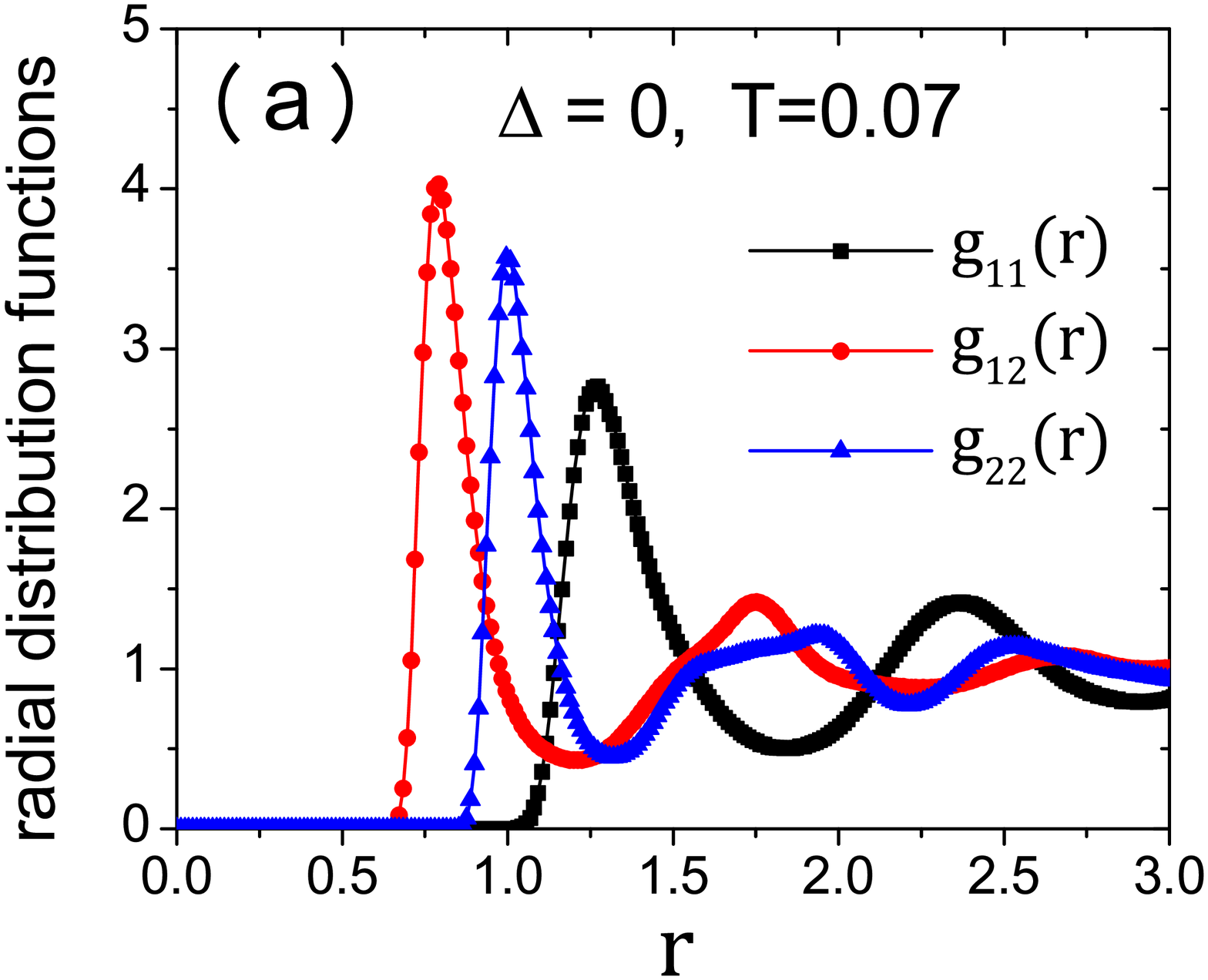}
\includegraphics[width=0.48\columnwidth]{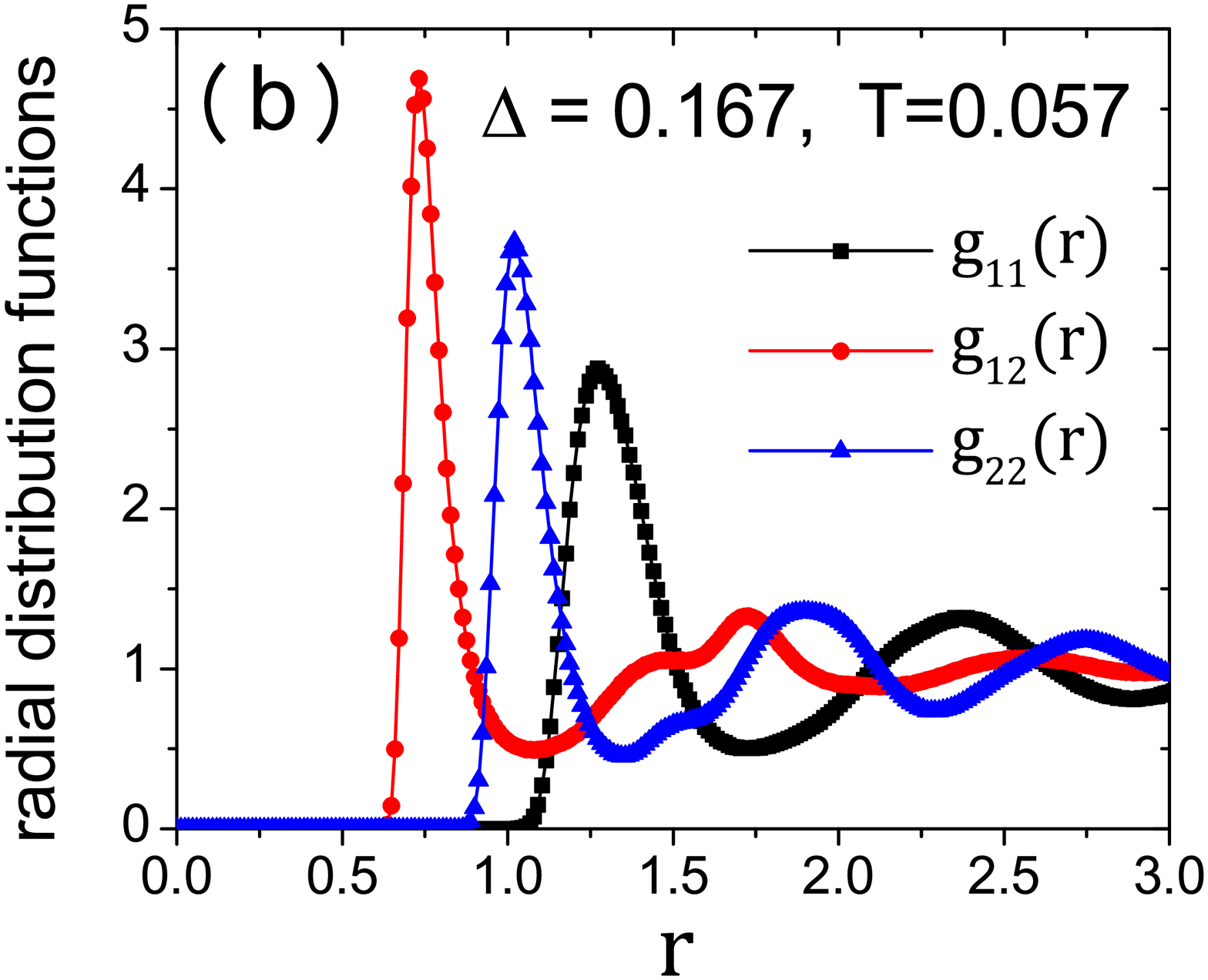}
\includegraphics[width=0.48\columnwidth]{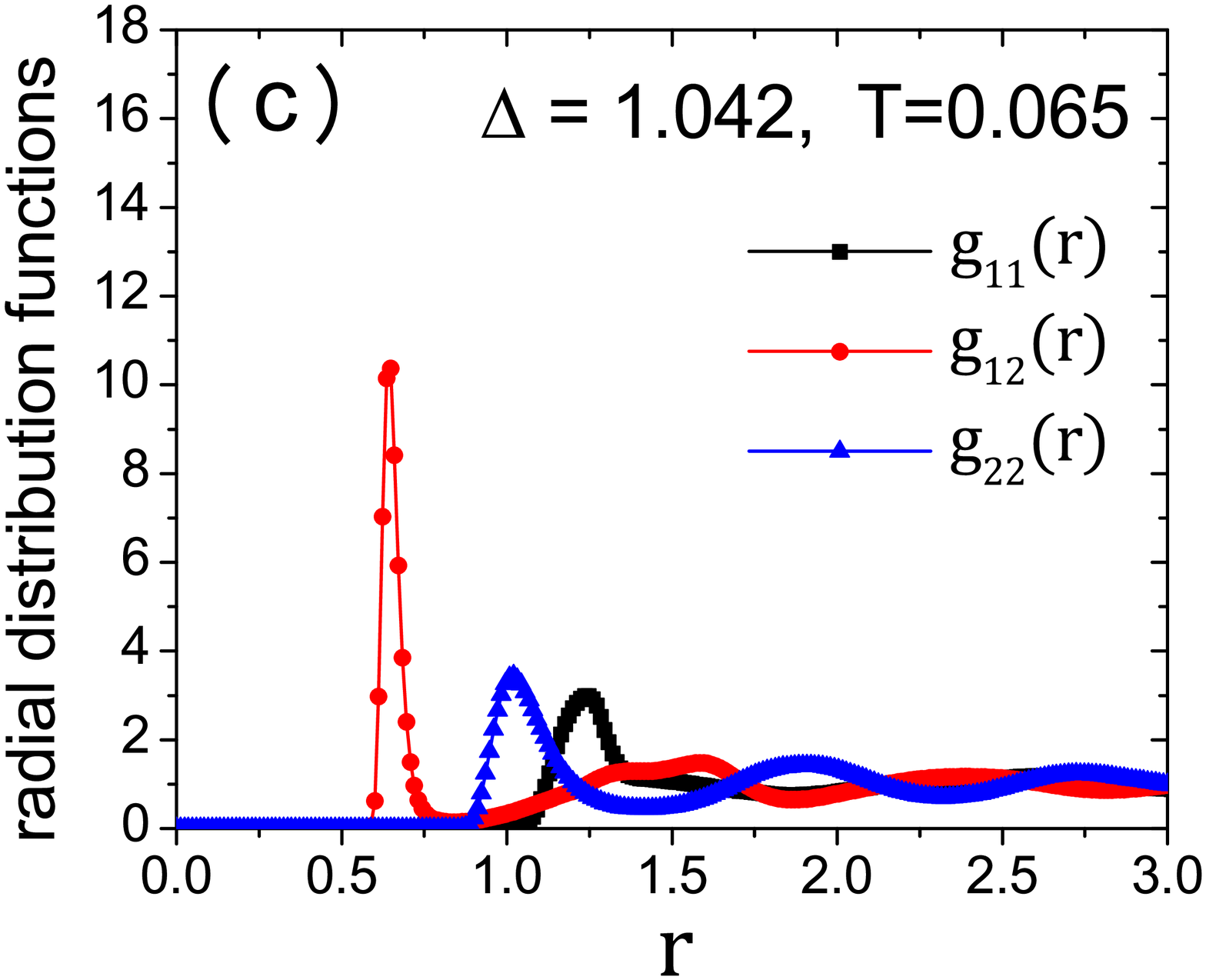}
\includegraphics[width=0.48\columnwidth]{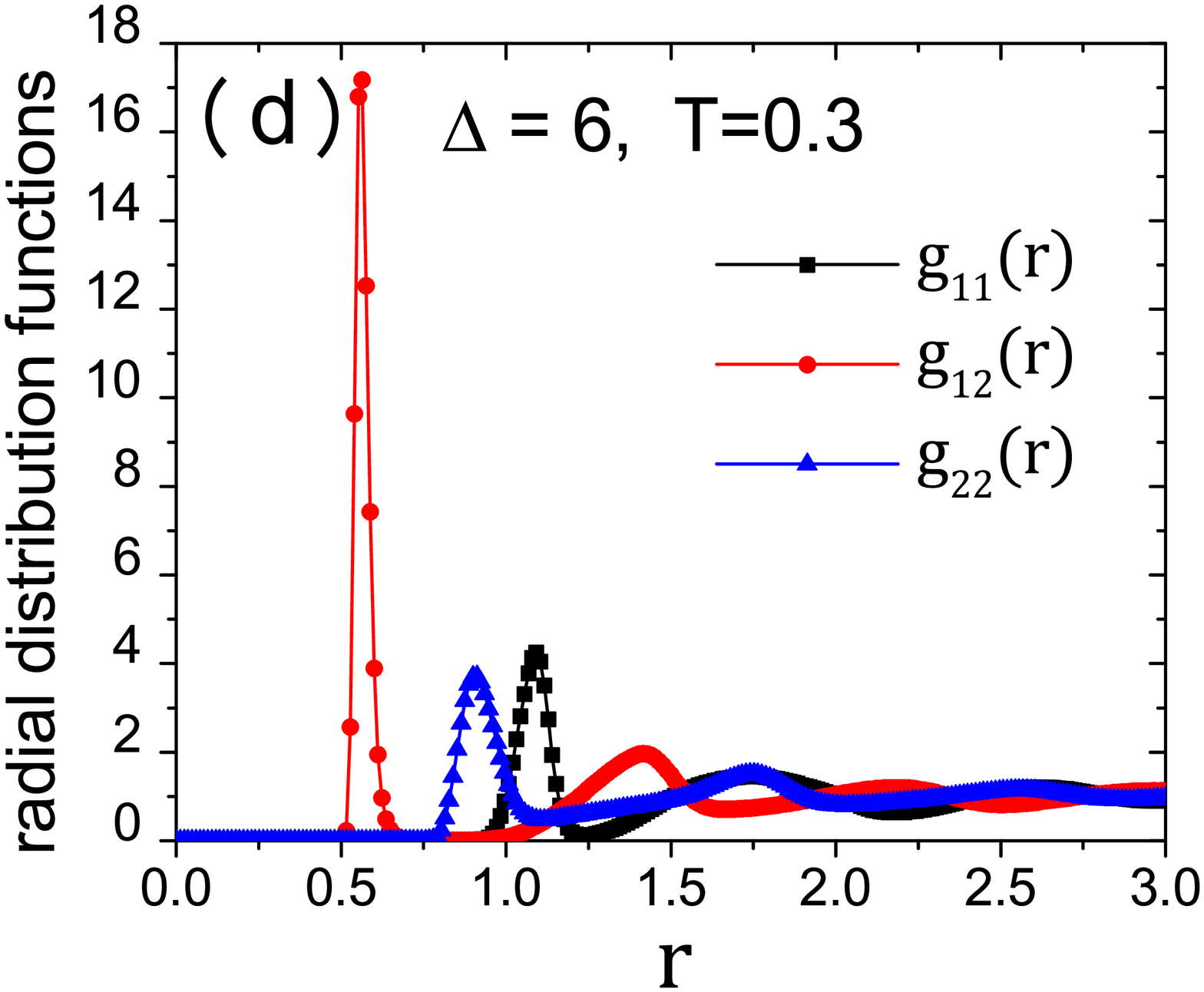}
\includegraphics[width=0.48\columnwidth]{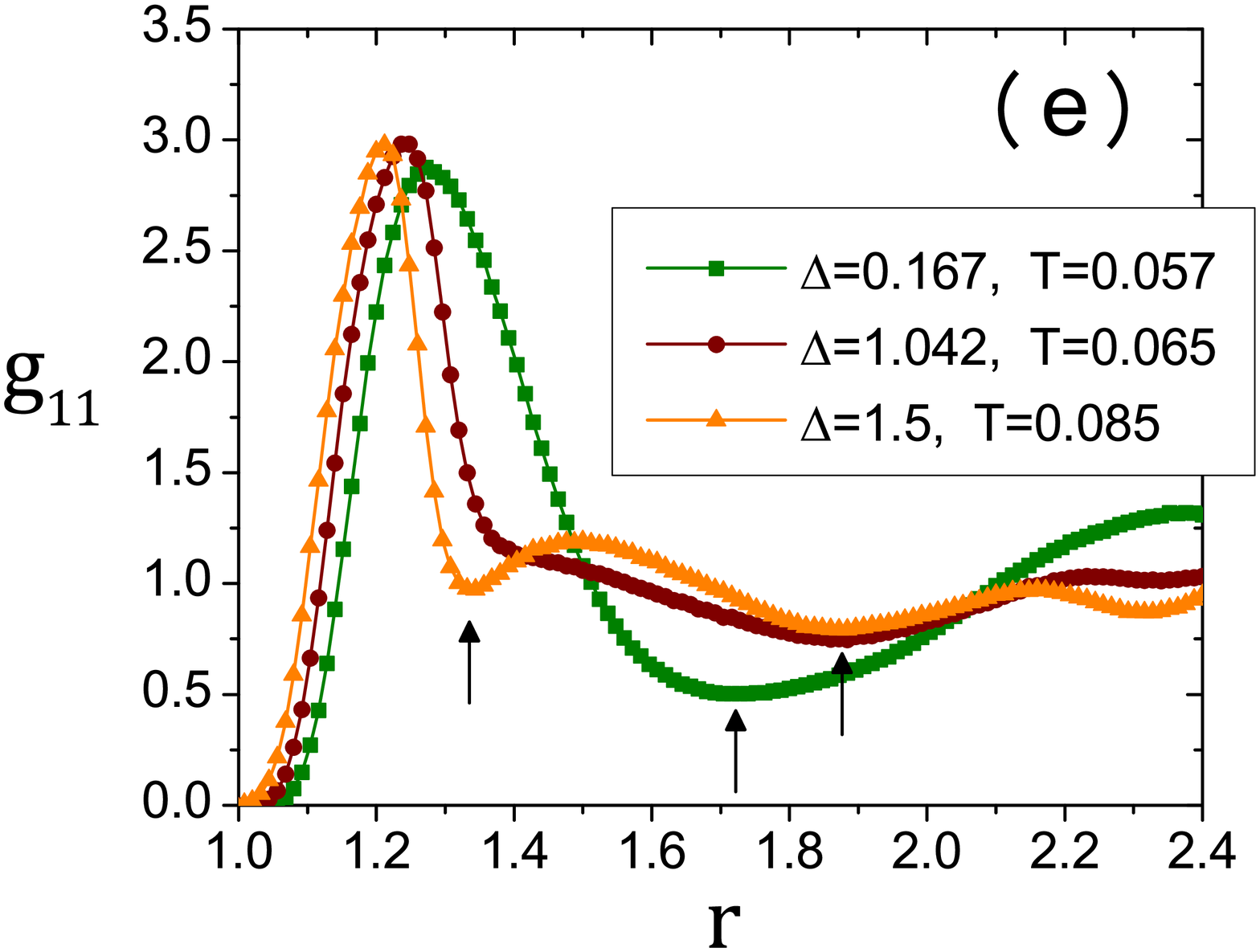}
\includegraphics[width=0.48\columnwidth]{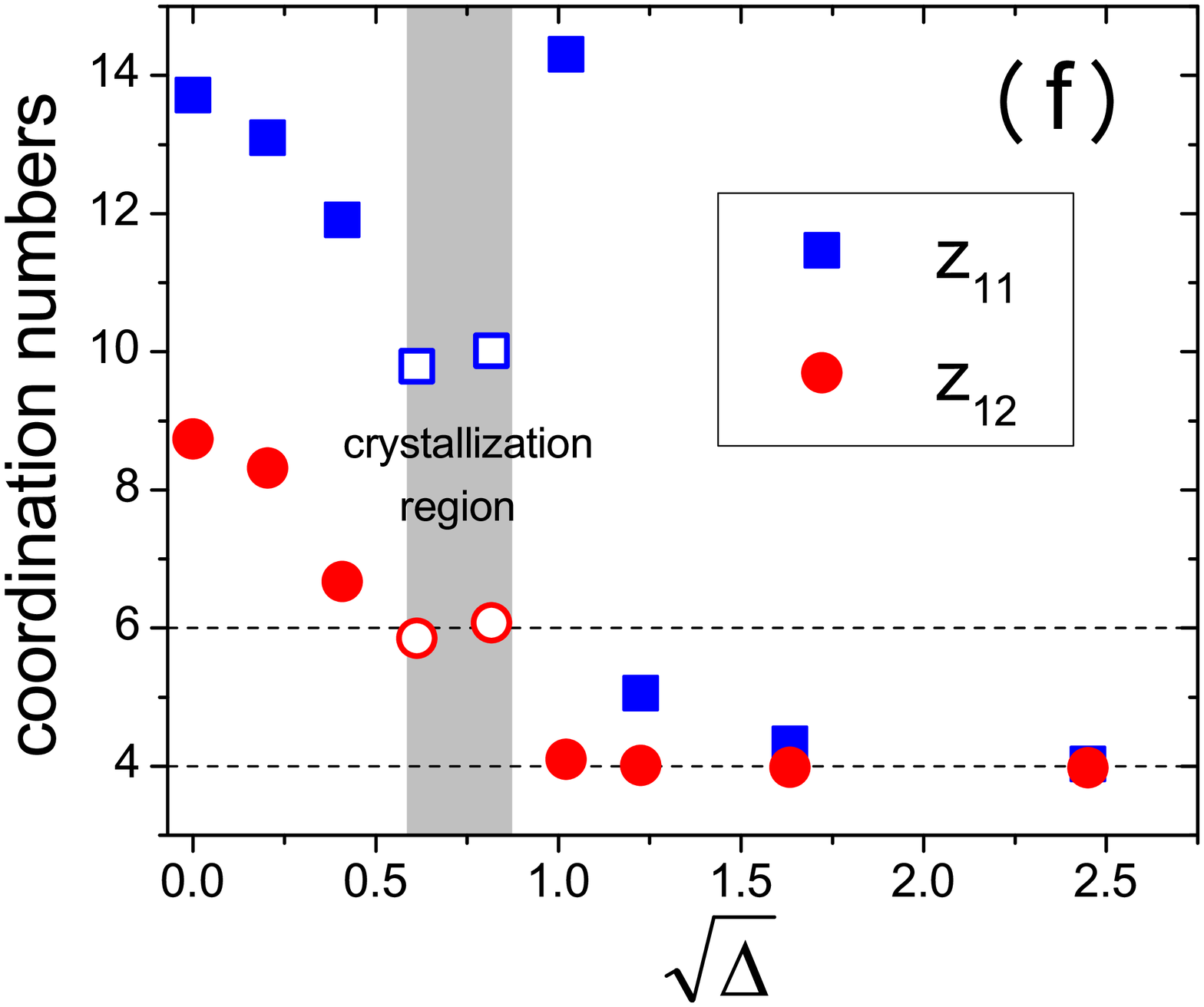}
\caption{
(a)--(d): The radial distribution functions, $g_{11}(r)$, $g_{12}(r)$, and $g_{22}(r)$, for several $\Delta$'s.
The data for the lowest temperature for each $\Delta$ are shown.
(e): $g_{11}(r)$ for several $\Delta$'s around 1. 
The vertical arrows are the position $r_{11}^*$ of the first minimum in $g_{11}(r)$.
(f): The coordination numbers $z_{11}$ and $z_{12}$ as a function of $\sqrt{\Delta}$.
Data for the lowest temperatures are shown for each $\Delta$.
The open symbols are the results for the crystalline structure. 
}
\label{fig:gofr}
\end{center}
\end{figure}

We first study the structural properties of the CP model for various
potential depth $\Delta$. We start with the radial distribution function
$g_{\alpha \beta}(r)$.
Figure~\ref{fig:gofr} shows $g_{\alpha \beta}(r)$ for each $\Delta$
at the lowest temperatures in our simulation which are listed in Table~\ref{tab:parameters}.
For $\Delta=6$ (see Figure~\ref{fig:gofr} (d)), $g_{12}(r)$ exhibits a very
sharp peak at $r \simeq 0.6\sigma_{11}$ followed by the very broad and small first minimum which persists up to
$r \simeq \sigma_{11}$ beyond which the small and broad second peak appears.
In contrast, the first peaks of $g_{11}(r)$ and $g_{22}(r)$ are observed 
at $r \simeq \sigma_{11}$ and $r \simeq \sigma_{22}$, respectively.
This indicates that the particles of species 1 and 2 are strongly
bonded~\cite{coslovich2009dynamics}. 
As $\Delta$ decreases, the first peak of $g_{12}(r)$ gradually
decreases and broadens.
Concomitantly the height of the first minimum rises to a
finite value at around $r\simeq \sigma_{11}$. 
Finally, at $\Delta=0$  (see Figure~\ref{fig:gofr} (a)), the profiles of
$g_{11}(r)$, $g_{12}(r)$, and $g_{22}(r)$ become similar to those of
typical fragile glass formers such as the SS and LJ binary mixtures~\cite{kob1995testing1,bernu1987soft}.
This structural change reflects the fact that the bonds between species 1 and 2 are weakened by decreasing $\Delta$.


To understand the effect of $\Delta$ on the structure in more detail, we
calculate the coordination number $z_{\alpha \beta}$. 
$z_{\alpha \beta}$ is the averaged number of the $\beta$ particles  in the
first neighbor shell of the $\alpha$ particle and defined by
\begin{equation}
z_{\alpha \beta}= \rho_{\beta} \int_0^{r_{\alpha \beta}^*} dr 4\pi r^2
 g_{\alpha \beta}(r),
\end{equation} 
where $\rho_{\beta}$ is the number density of species $\beta$ and
$r_{\alpha \beta}^*$ is the position of the first minimum of $g_{\alpha
\beta}(r)$.
In particular, $z_{11}$ and $z_{12}$
are useful to characterize the
tetrahedral network structure.
In $\rm SiO_2$, a single Si atom (species 1) is surrounded by
four oxygen atoms (species 2), thus, 
$z_{11}=4$, $z_{12}=4$, and $z_{21}=2$. 
In Figure~\ref{fig:gofr} (f), we show $z_{11}$ and $z_{12}$ at the lowest 
temperatures
for several $\Delta$'s.
We use $\sqrt{\Delta}$ instead of $\Delta$ as the horizontal axis for the sake
of visual clarity.
At $\Delta=6$ ($\sqrt{\Delta} \simeq 2.449$), we obtain $z_{11} = 4$ and $z_{12}=4$, corresponding to the perfect tetrahedral
network structure~\cite{coslovich2009dynamics}. 
As $\Delta$ decreases, both $z_{11}$ and $z_{12}$ increase. 
Eventually, we observe $z_{11} \simeq 14$ and $z_{12} \simeq 9$ at $\Delta=0$.
This indicates that the tetrahedral network structure is broken and that
the local configuration of the system becomes more isotropic and compact.
We find that, in a narrow range of $\Delta$, $0.375 \leq  \Delta \leq 0.667$  (or $0.612 \lesssim \sqrt{\Delta} \lesssim  0.816$),
the system crystallizes at low temperatures. 
Some of the samples in our simulation runs also crystallized at the lowest $T$ at $\Delta=0.167$ and $1.042$, 
when we carried out simulation for very long time ($t \gtrsim 10^6$).
We show results for the crystalline state by open
symbols in Figure~\ref{fig:gofr}~(f).
We confirm by carefully inspecting the real space snapshots that the systems crystallize
completely over the entire simulation box.
The coordination numbers for crystallized samples are $z_{12} = 6$ and $z_{11} = 10$.
This suggests that the observed crystalline structure is Stishovite type.
Indeed, it has been reported that $\rm SiO_2$ forms the Stishovite crystal under very high pressures
~\cite{keskar1992structural}.
We remark that an exceptionally large value of $z_{11}$ 
$\gtrsim 14$
at $\Delta=1.042$ ($\sqrt{\Delta} \simeq 1.021$) is due to the ambiguity to define the first coordination shell.
As shown in Figure~\ref{fig:gofr}~(e), the profile of $g_{11}(r)$ is broadened
around $\Delta \simeq 1$, which makes it difficult to define $r_{11}^*$ clearly.
Thus, $\Delta \simeq 1$ can be regarded as the crossover regime from the
network dominated region to the isotropic/compact structure region.

\begin{figure}
\begin{center}
\includegraphics[width=0.48\columnwidth]{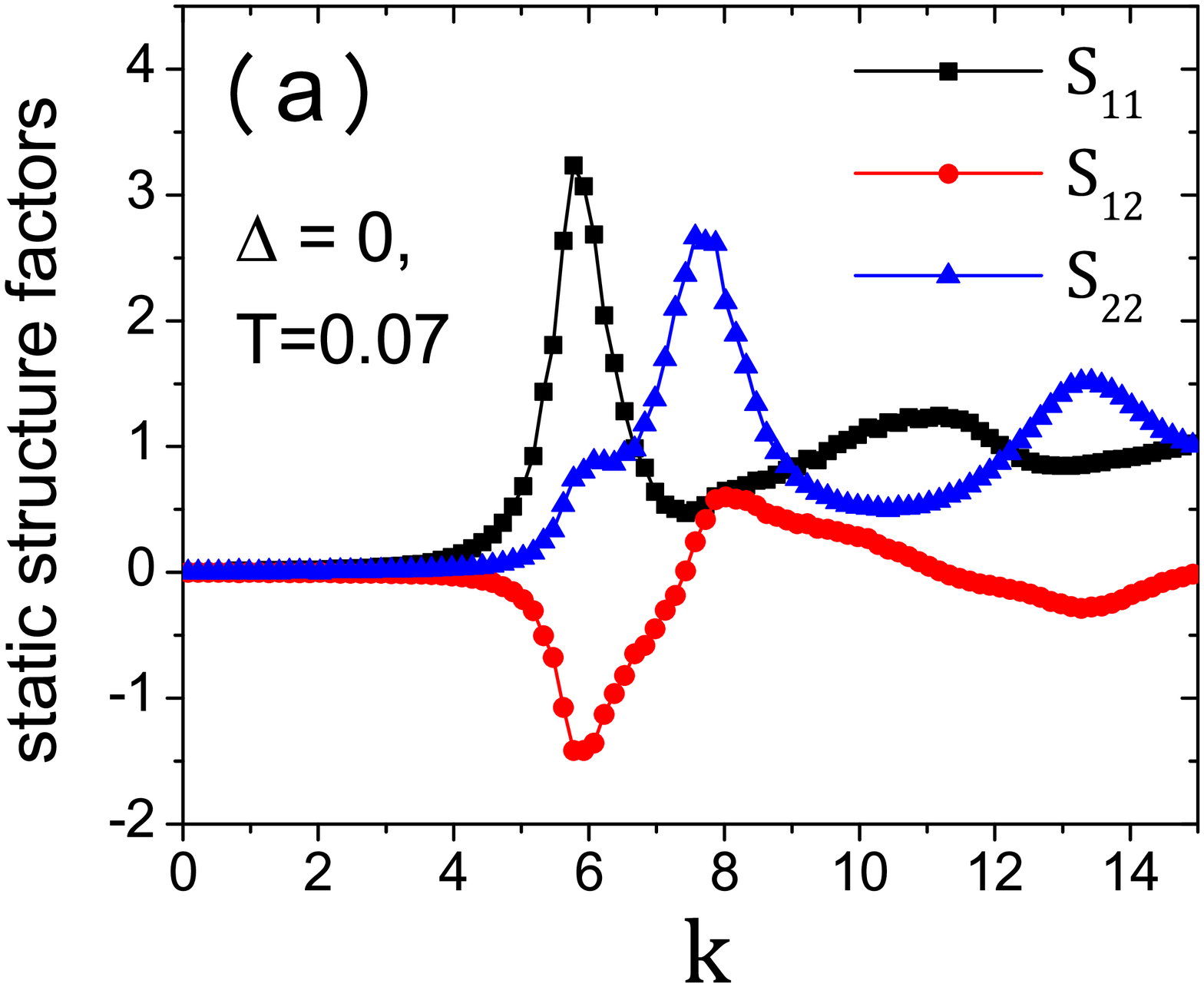}
\includegraphics[width=0.48\columnwidth]{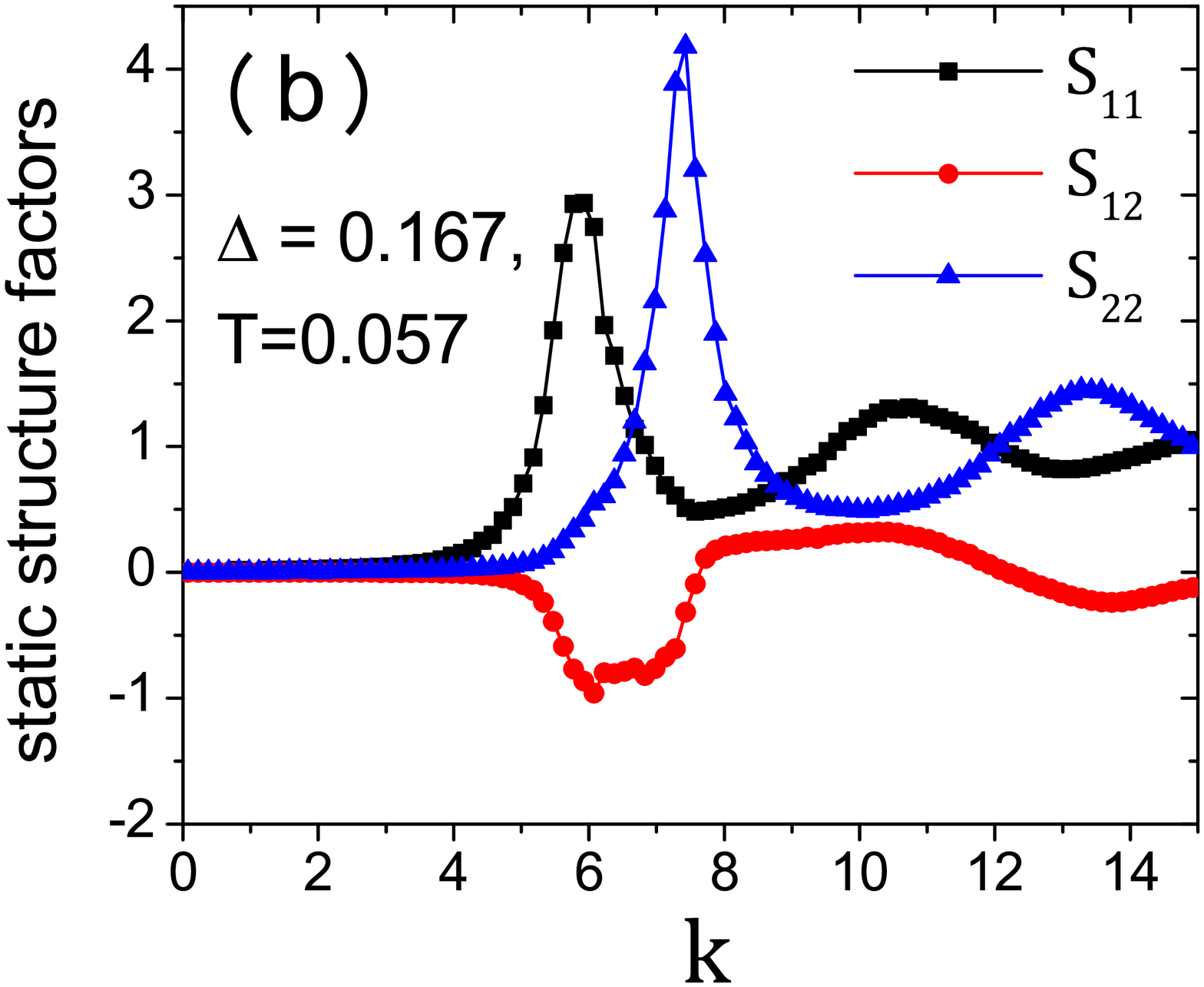}
\includegraphics[width=0.48\columnwidth]{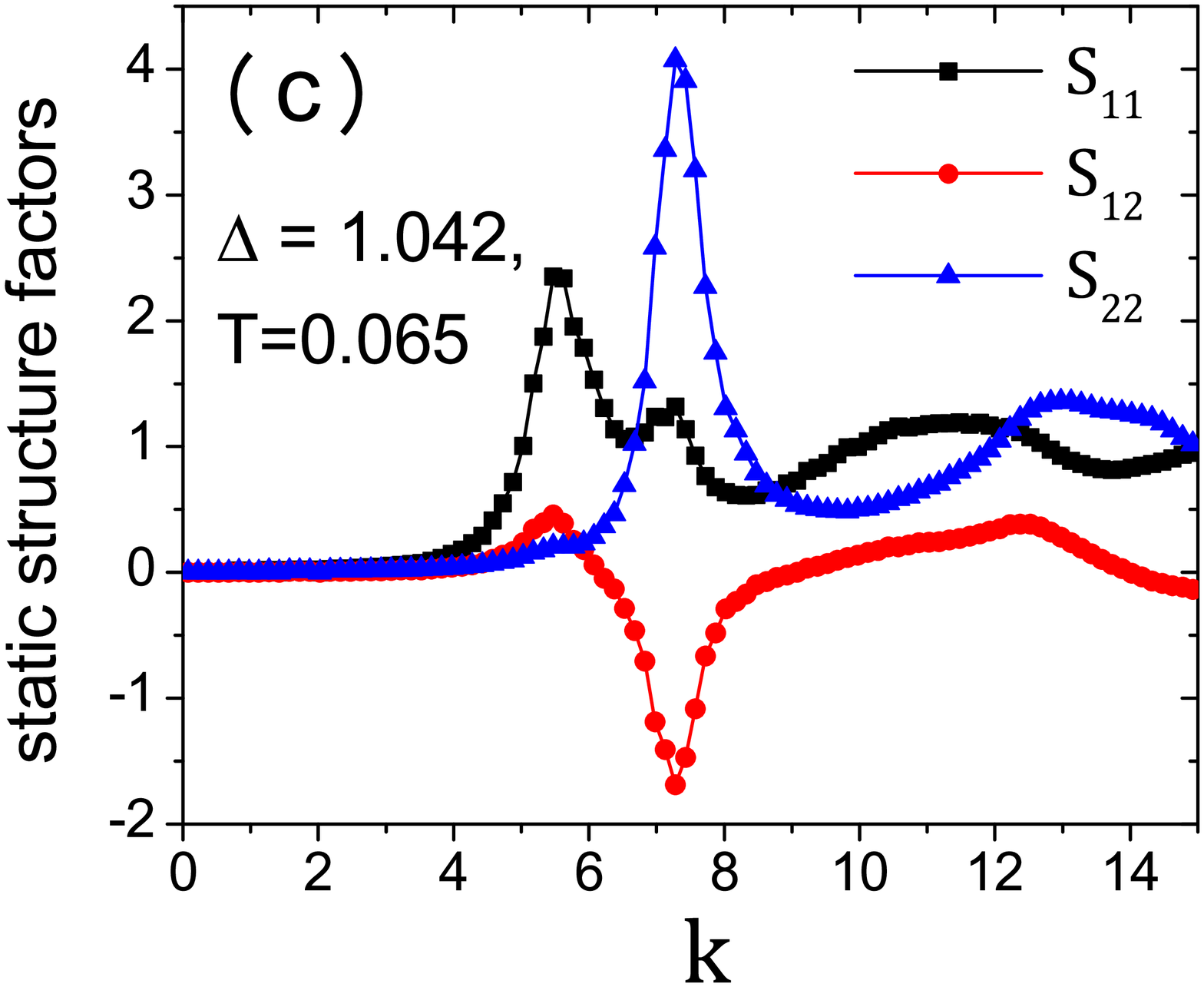}
\includegraphics[width=0.48\columnwidth]{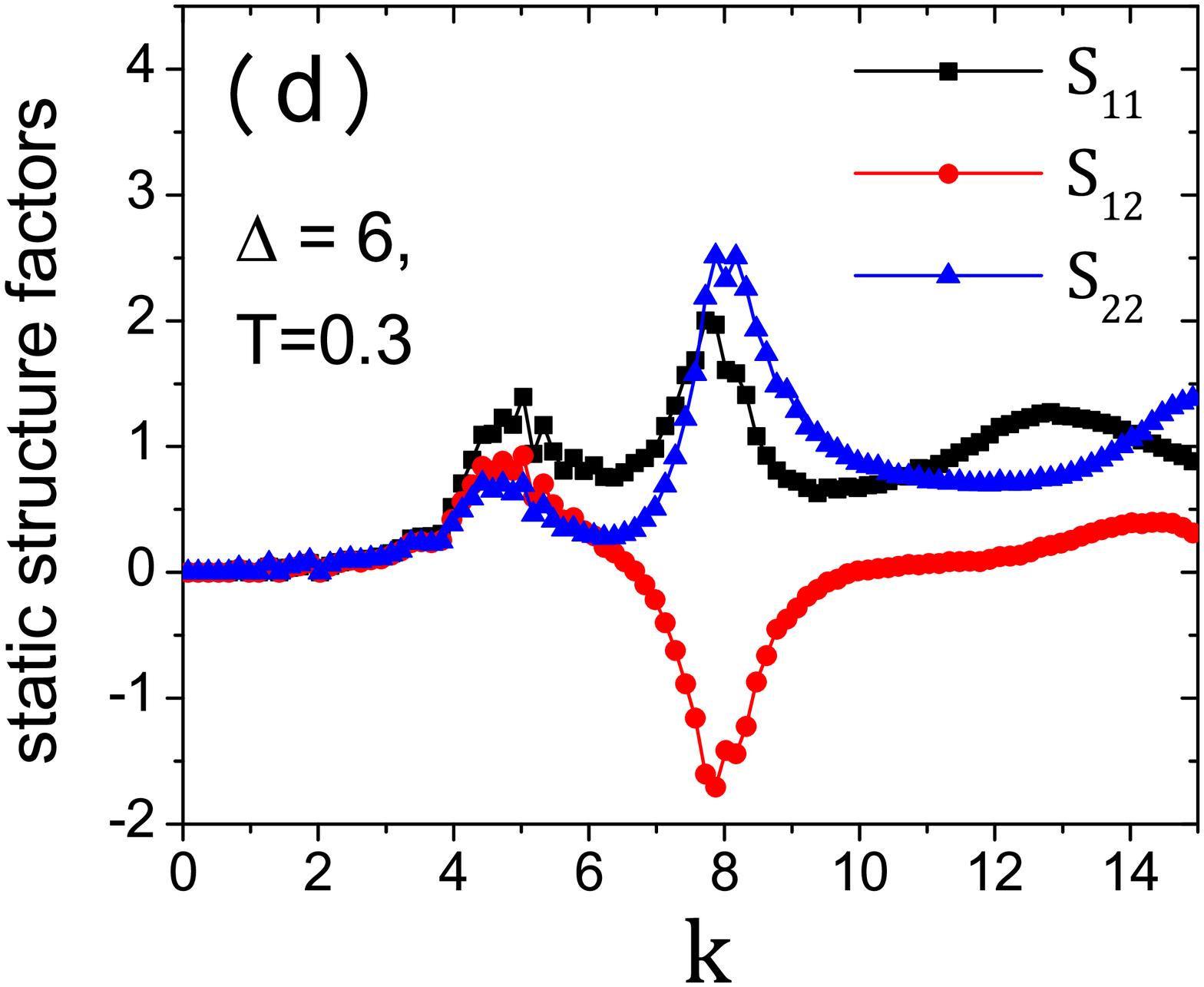}
\caption{
(a)--(d): The static structure factors, $S_{11}(k)$, $S_{12}(k)$, and $S_{22}(k)$, for several $\Delta$'s.
The results for the lowest temperatures of our simulations are shown.
}
\label{fig:Sofk}
\end{center}
\end{figure}

Next, we calculate the static structure factor $S_{\alpha \beta}(k)$.
We show $S_{\alpha \beta}(k)$ at the lowest temperatures in
Figure~\ref{fig:Sofk} for several $\Delta$'s.
At $\Delta=6$, the maximal peaks are observed around $k \simeq 8$ in all 
components of $S_{\alpha \beta}(k)$.
These peaks originate from the tetrahedral structure whose bond length
is estimated to be $2\pi /k \simeq 0.79$.
We also find weaker peaks at smaller wave numbers around $k \simeq 5$.
This is the so-called first sharp diffraction peak
(FSDP)~\cite{elliott1991origin}, which is known as an indication of
the medium-range order of the length scale larger than the neighbor shell. 
In particular, the FSDP has been attributed to the formation of the hierarchical clusters 
of the tetrahedra with larger lengths~\cite{nakamura2015description}.
When the potential depth $\Delta$ is reduced, these two peaks merge
to a single peak.
This result implies that the hierarchical tetrahedral structures disappear
gradually as $\Delta$ is decreased.
At $\Delta \lesssim 1$, the shape of the static structure factors are analogous to
those observed in typical fragile glass formers such as the SS and LJ binary mixtures~\cite{kob1995testing2,kob2012spatial}.

\begin{table} 
\begin{center} 
\scalebox{0.85}{
\begin{tabular}{cc|ccccccccc}
\hline
\hline
\ $\Delta$  & \ $\sqrt{\Delta}$ \ & \ $\rho$ \ & \ $T$ \ & \ $k^*$ \ & \ $T_{\rm onset}$ \ & \ $T_0^{(\tau_{\alpha})}$  & \ $T_0^{(D)}$ \ & \ $T^*$ \ & \ $K^{(\tau_{\alpha})}$ \ & \ $K^{(D)}$ \ \\
\hline
0  & \ 0 \ & \ $\infty$ \ & 0.07-0.34 \ &  5.8 \ & 0.18 \ & 0.0548 \ & 0.0547 \ & 0.085 \ & 0.372 \ & 0.475 \ \\
0.042  & \ 0.204 \ & \ 5.733 \ & 0.06-0.34 \ &  5.9 \ & 0.16 \ &  0.0467 \ & 0.0465 \ & 0.065 \ & 0.357 \ & 0.459 \  \\
0.167  & \ 0.408 \  & \ 4.054 \ & 0.057-0.34 \ & 5.9 \ & 0.12 \ & 0.0472 \ & 0.0461 \ & \textrm{Non} \ & 0.652 \ & 0.759 \ \\
1.042  & \ 1.021 \  & \ 2.564 \ & 0.065-0.34 \ &  5.6 \ & 0.14 \ & 0.046 \ & 0.0456 \ & 0.075 \ & 0.241 \ & 0.286 \ \\
1.5  & \ 1.225 \  & \ 2.341 \ & 0.085-0.34 \ &  5.5 \ & 0.18 \ & 0.0599 \ & 0.059 \ & 0.11 \ & 0.2 \ & 0.229 \ \\
2.667 & \ 1.633 \  & \ 2.027 \ & 0.14-0.6 \ & 5.2 \ & 0.26 \ & 0.0776 \ & 0.0793 \ & 0.2 \ & 0.1 \ & 0.126 \ \\
6  & \ 2.449 \  & \ 1.655 \ & 0.3-1.25 \ & 5.0 \ & 0.48 \ & 0.161 \ & 0.173 \ & 0.4 \ & 0.087 \ & 0.123 \ \\
\hline \hline
\end{tabular}
}
\caption{
Summary of the parameters, $\Delta$, the corresponding densities $\rho$, and the temperature
 range of the present study.
The peak position $k^*$ of $S_{11}(k)$, the onset temperature $T_{\rm onset}$ of the
 two-step relaxation, $T_0$ obtained by fitting of the VFT equations from the relaxation time
 $\tau_{\alpha}$ and diffusion constant $D$, the peak position $T^*$ of the specific heat, and
the fragility index $K$ from $\tau_{\alpha}$ and $D$ are also tabulated.
} 
\label{tab:parameters}
\end{center}
\end{table}

\subsection{Dynamical properties}

As demonstrated above, 
the tetrahedral network structures are broken when the potential depth
$\Delta$ is reduced to zero.
In this section, we analyze the dynamical properties of the CP model.
Specifically, the $\Delta$ dependence of the fragility is quantified.

\begin{figure}
\begin{center}
\includegraphics[width=0.48\columnwidth]{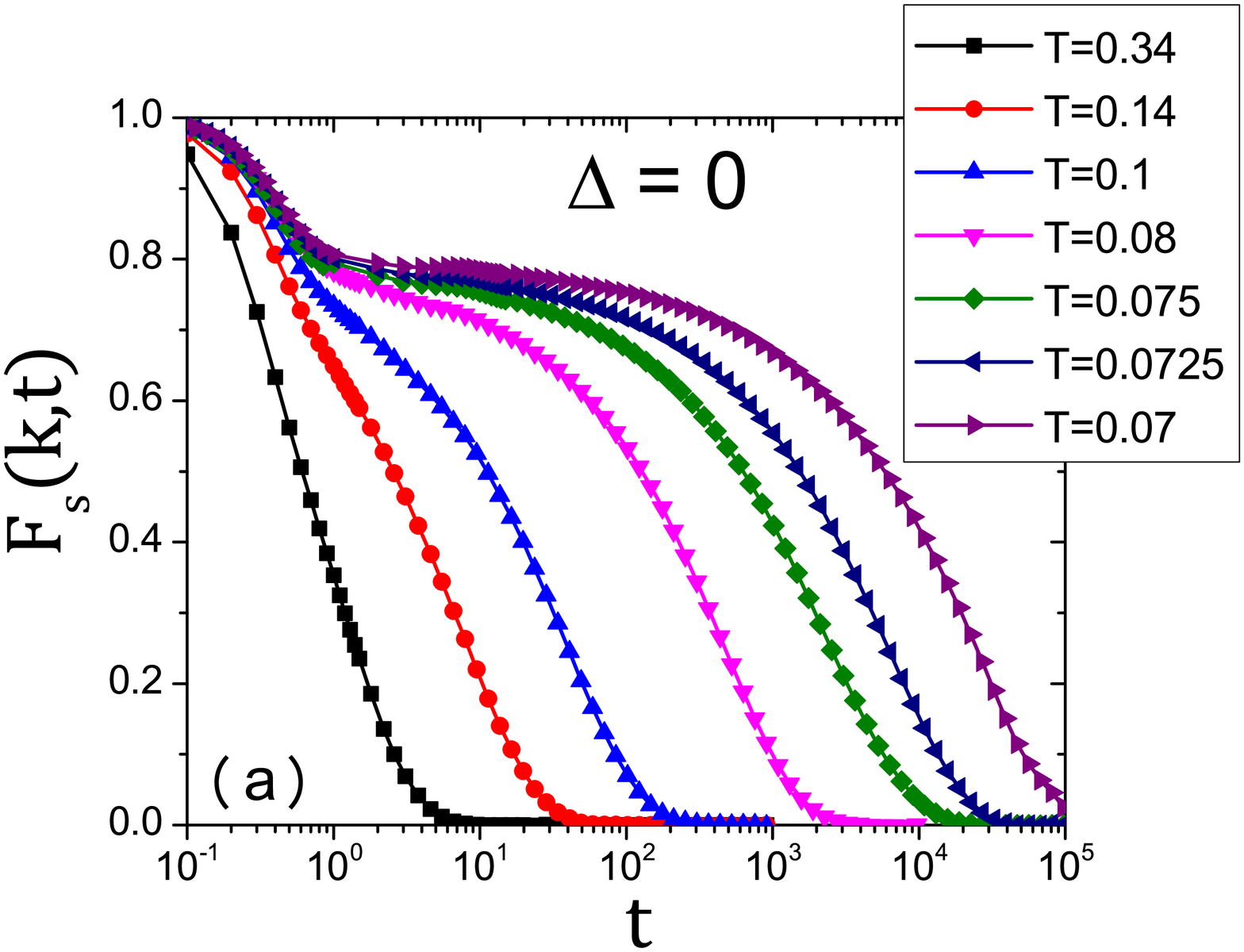}
\includegraphics[width=0.48\columnwidth]{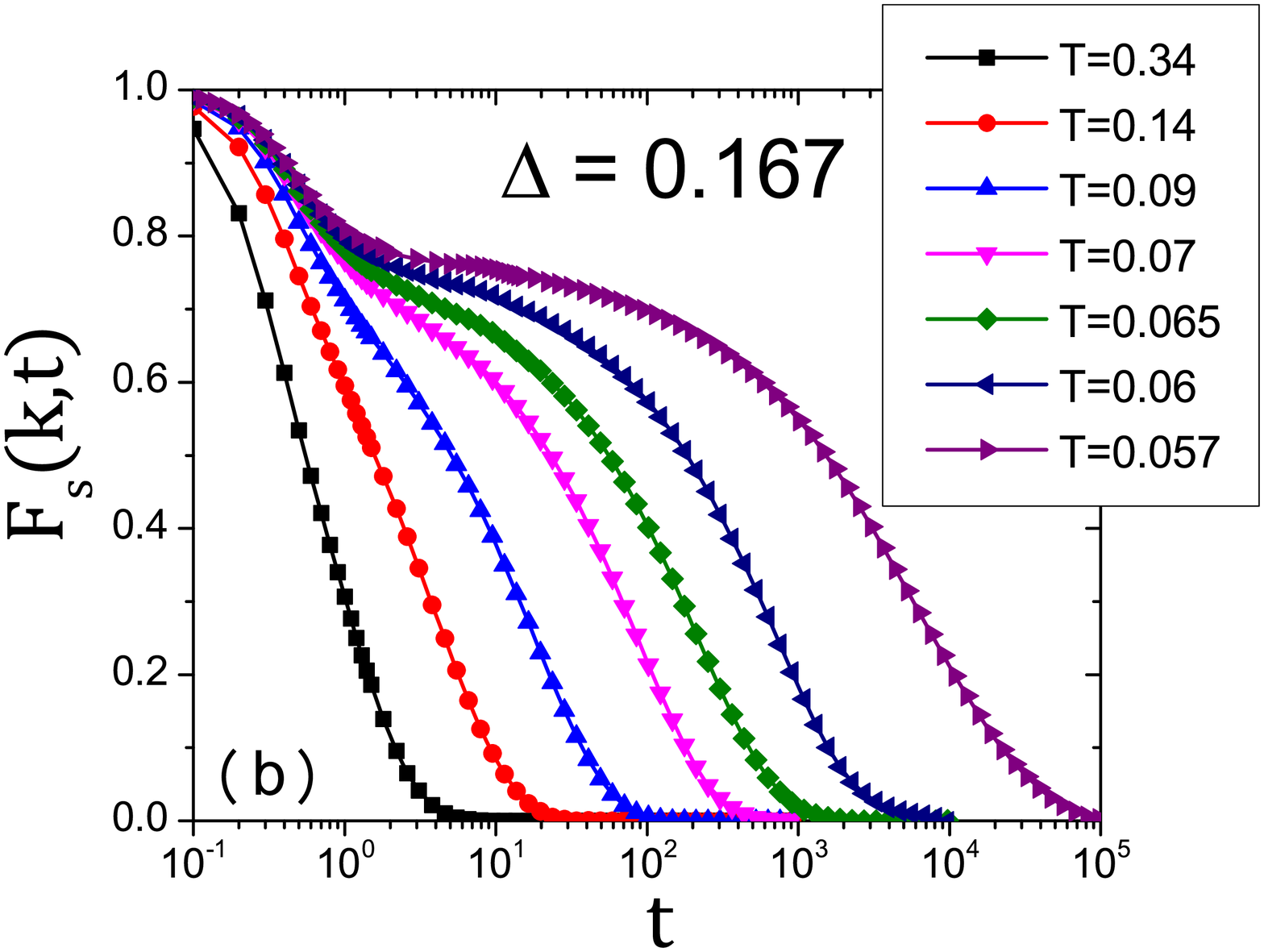}
\includegraphics[width=0.48\columnwidth]{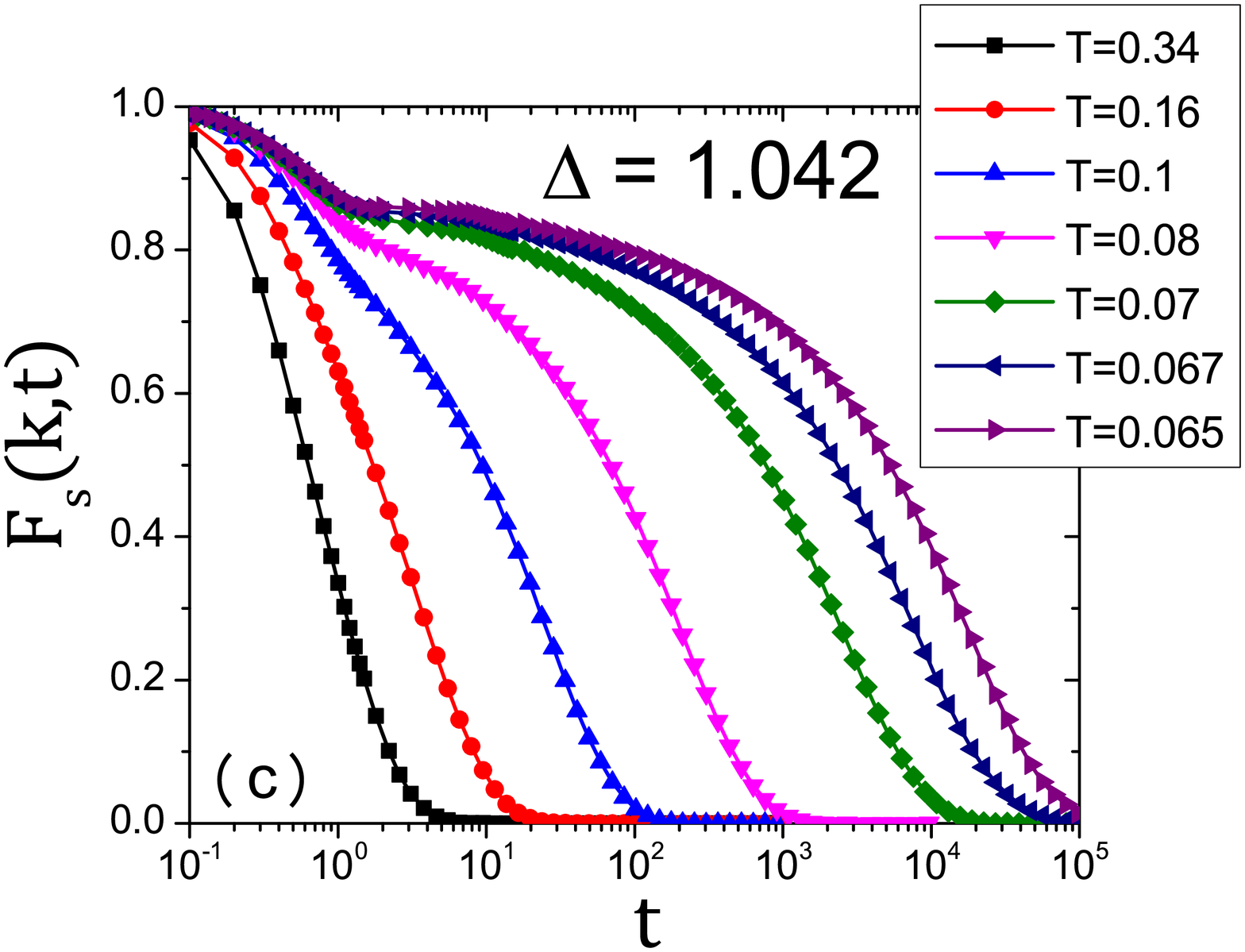}
\includegraphics[width=0.48\columnwidth]{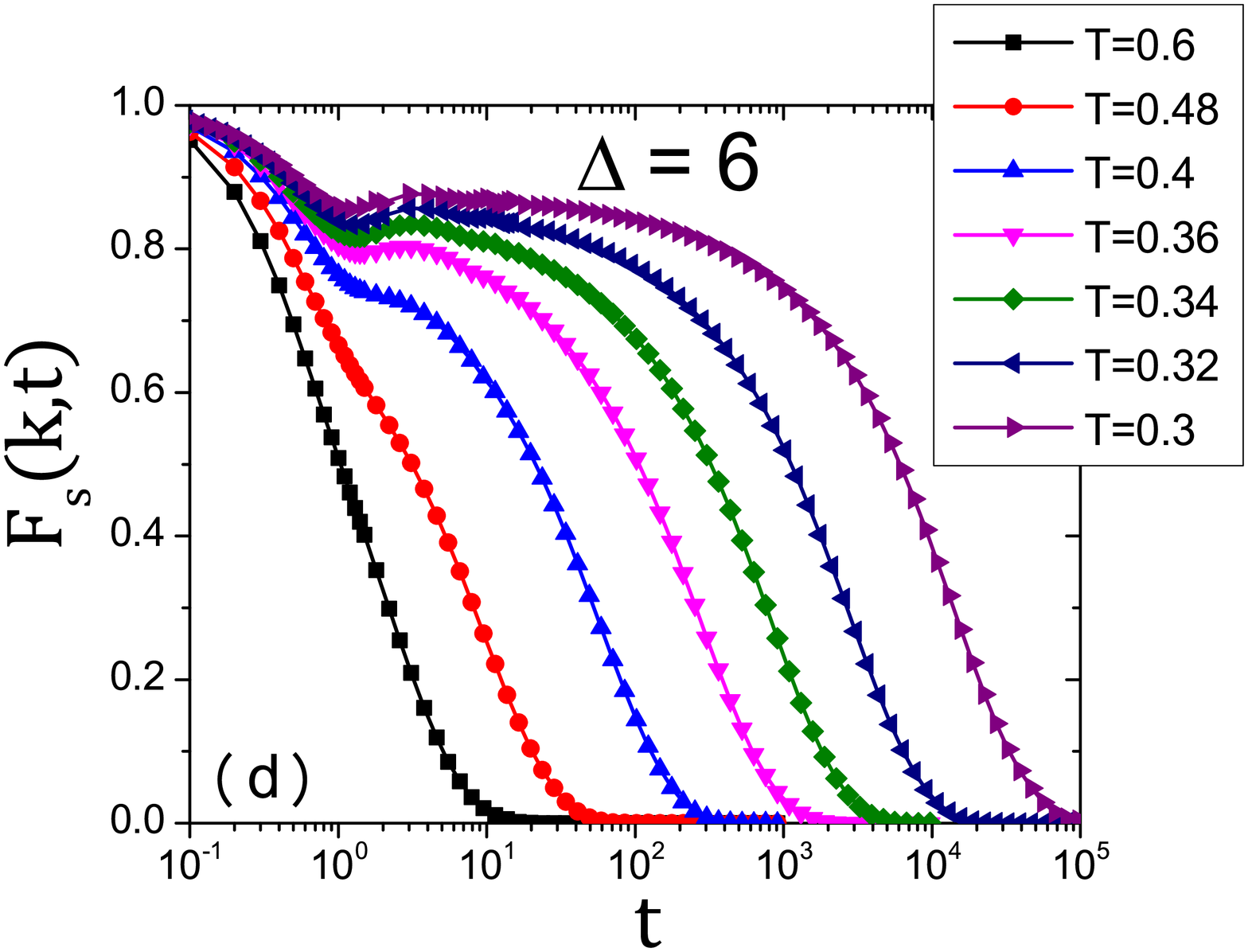}
\caption{
(a)--(d): The temperature variation of the self part of the intermediate scattering function for species 1 for several $\Delta$'s.
}
\label{fig:Fskt}
\end{center}
\end{figure}  

First, we calculate the self part of the intermediate
scattering function $F_s(k, t)$ for species 1 defined by
\begin{equation}
F_s(k, t) = \frac{1}{N_1} \sum_{i=1}^{N_1} \left\langle e^{-i {\bm
					    k}\cdot \left({\bm r}_i(t) -
						     {\bm
						     r}_i(0)\right)}
					   \right\rangle,
\label{eq:Fskt}
\end{equation}
where ${\bm r}_i(t)$ denotes the position of the $i$-th particle at time $t$.
In Figure~\ref{fig:Fskt}, we show the temperature dependence of 
$F_s(k, t)$ for several $\Delta$'s.
The wave number $k=|\bm{k}|$ is chosen at the peak position $k^*$ of the
static structure factor, $S_{11}(k)$ (see Figure~\ref{fig:Sofk}).
The values of $k^*$ are listed in Table~\ref{tab:parameters}.
At high temperatures, $F_s(k, t)$ shows the exponential decay with the
short relaxation time but, at low temperatures, dynamics dramatically slows down and $F_s(k,t)$ exhibits a two-step relaxation.
This is the sign of the onset of the glassy dynamics.

\begin{figure}
\begin{center}
\includegraphics[width=0.48\columnwidth]{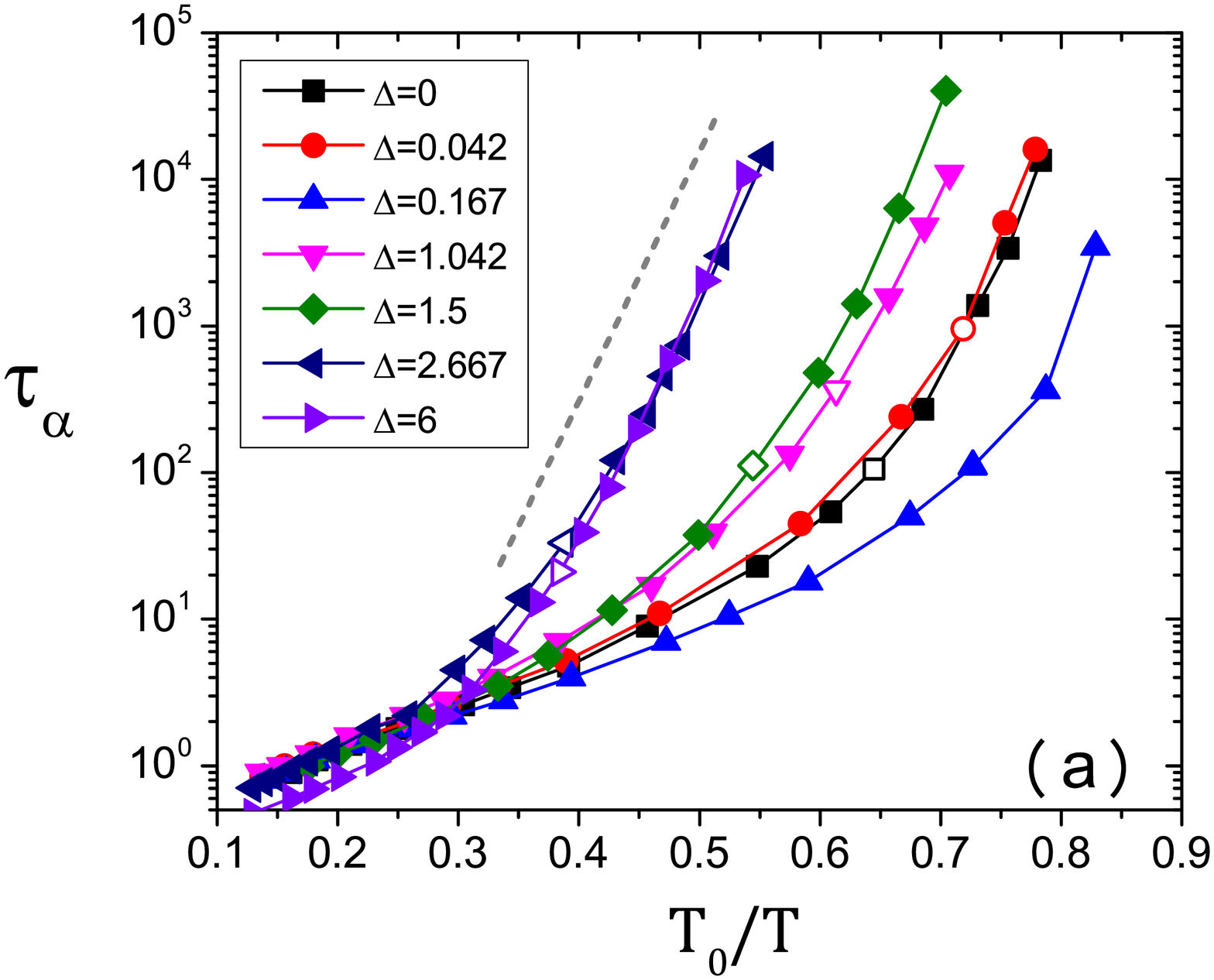}
\includegraphics[width=0.48\columnwidth]{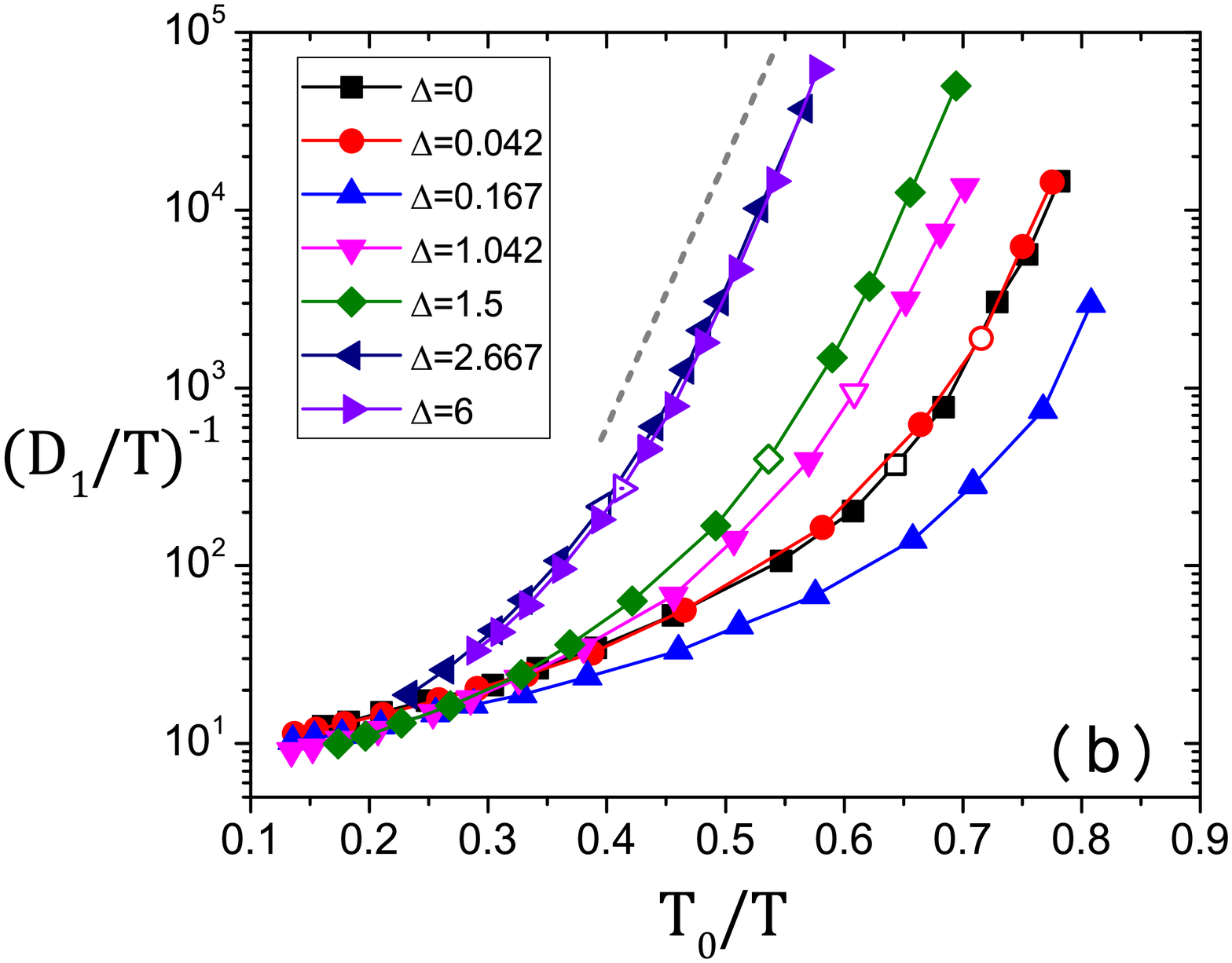}
\caption{
The Arrhenius plots of the relaxation time (a) and of the diffusion constant for
 species 1 (b)
as a function of the inverse temperature $1/T$.
In both plots, the temperature is scaled by $T_0$, a fitting parameter of the VFT equation.
The open symbols are the position of $T^*$ at which the specific heat shows the peak (see Figure \ref{fig:specific_heat}).
The dashed straight lines indicate the Arrhenius temperature dependence. 
}
\label{fig:Arrhenius_plot}
\end{center}
\end{figure}

We define the relaxation time $\tau_{\alpha}$ by $F_s(k, \tau_{\alpha}) = e^{-1}$. 
We fit the observed $\tau_{\alpha}$ by the Vogel-Fulcher-Tammann
(VFT) equation;
\begin{equation}
\tau_{\alpha} \sim \exp \left[ \frac{1}{K(T/T_0-1)} \right],
\label{eq:VFT}
\end{equation}
where $T_0$ and $K$ are fitting parameters.
The parameter $K$ is referred to as the fragility index which is regularly used to quantify the
degree of the super-Arrhenius temperature dependence~\cite{sastry2001relationship}.
The larger (smaller) values of $K$ correspond to the fragile (strong) glass formers.
In this study, the VFT fitting is applied below $T_{\rm onset}$, where
the two-step relaxation for $F_s(k,t)$ sets in.
$T_{\rm onset}$ for each $\Delta$ is presented in Table~\ref{tab:parameters}. 
We show the temperature dependence of $\tau_{\alpha}$ for several $\Delta$'s in
Figure~\ref{fig:Arrhenius_plot} (a).
The temperature $T$ is scaled by $T_0$.
It is clearly seen that the results for 
$\Delta=2.667$ and $\Delta=6$ follow the Arrhenius law at low temperatures~\cite{coslovich2009dynamics}.
For smaller $\Delta$'s, the temperature dependence of $\tau_{\alpha}$ deviates
from the Arrhenius behavior.
In other words, the system changes from strong to fragile glass formers
with decreasing $\Delta$.

\begin{figure}
\begin{center}
\includegraphics[width=0.48\columnwidth]{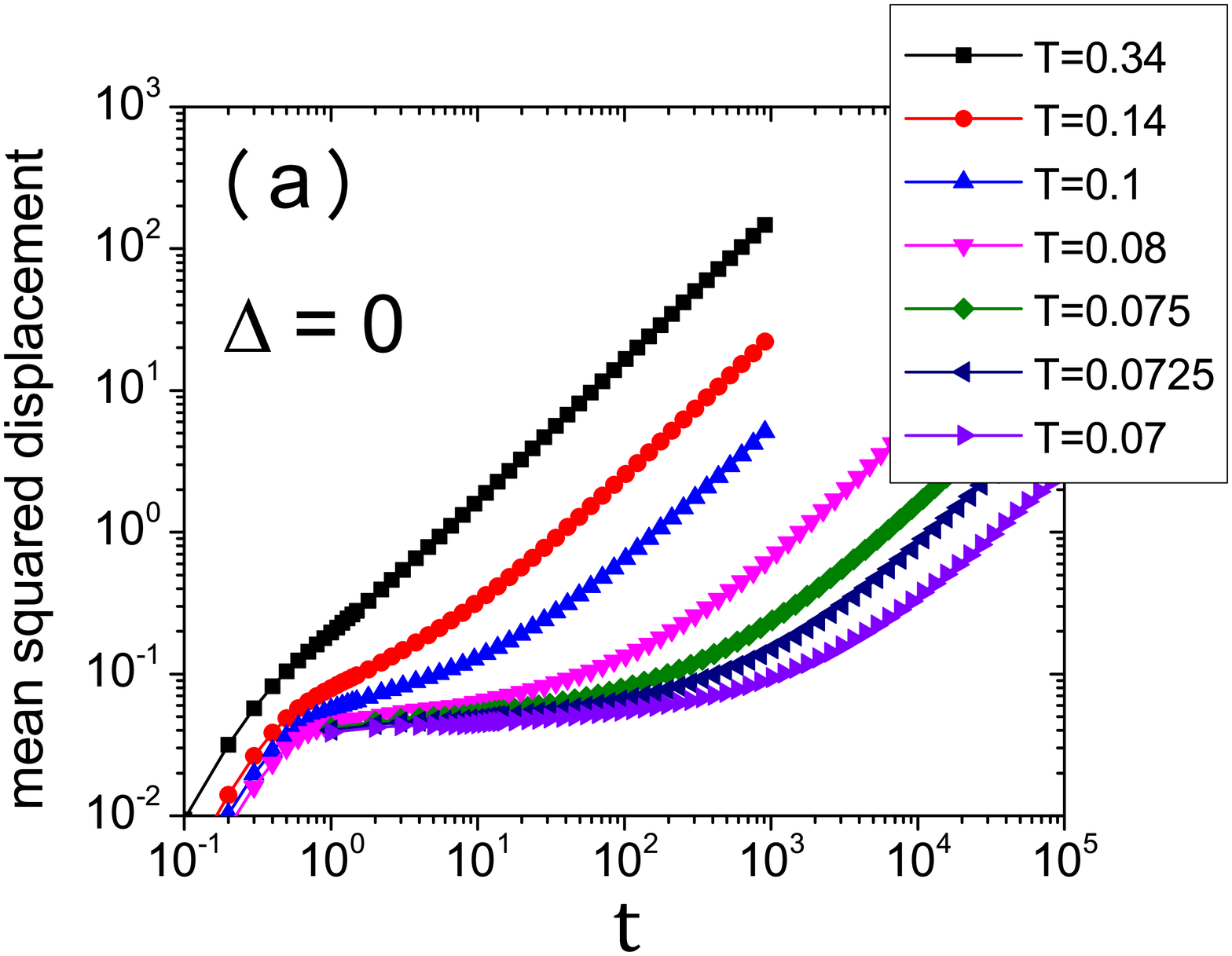}
\includegraphics[width=0.48\columnwidth]{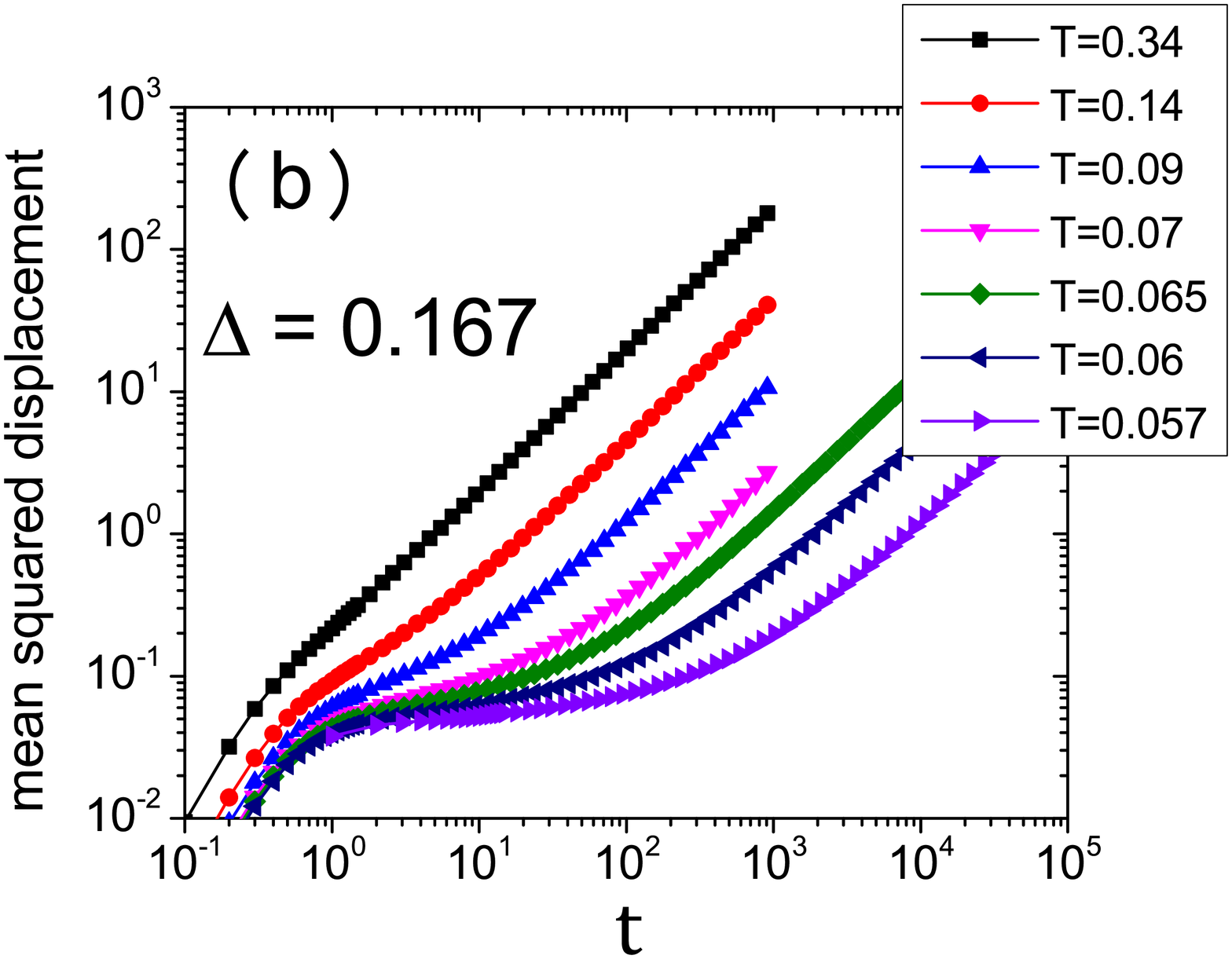}
\includegraphics[width=0.48\columnwidth]{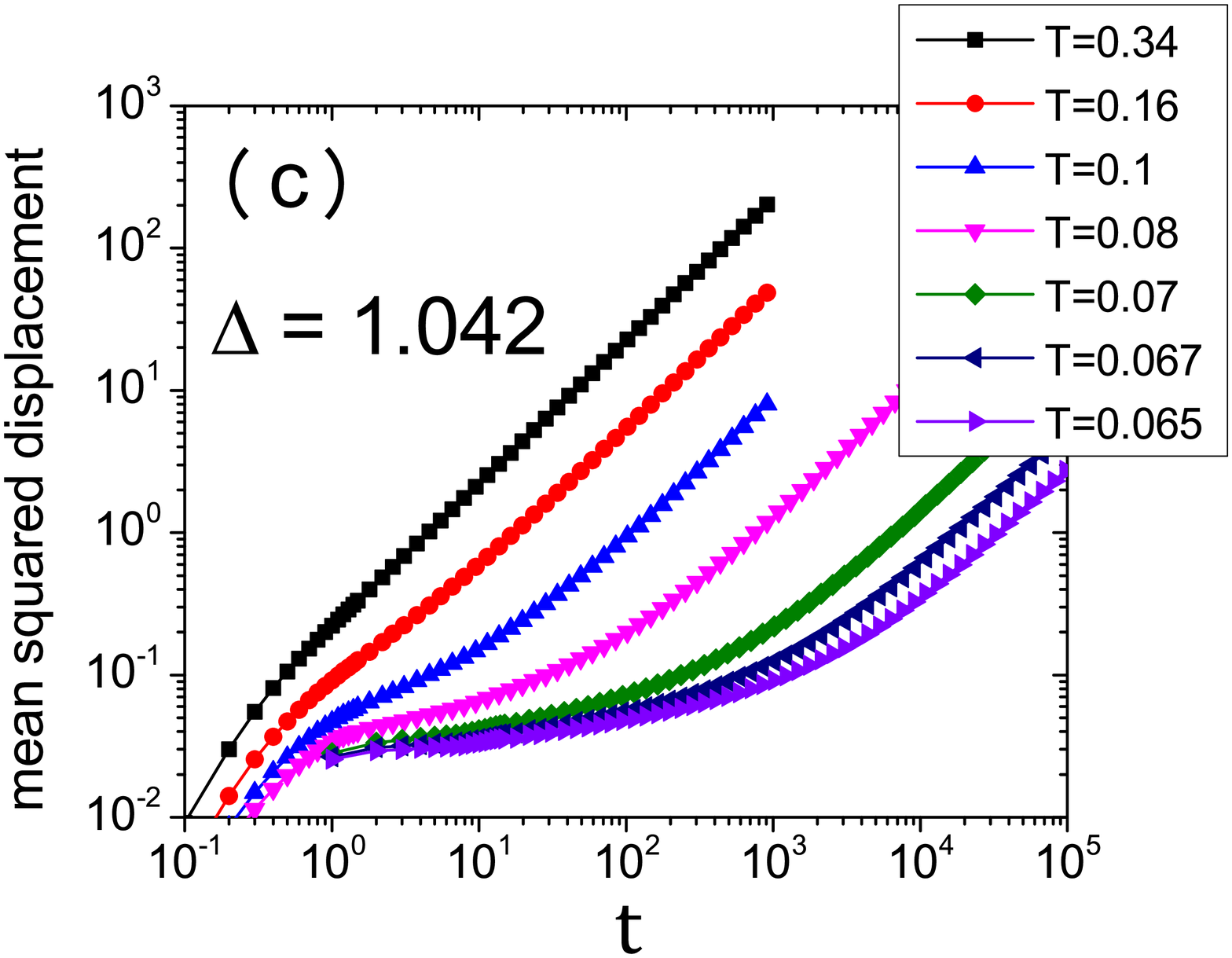}
\includegraphics[width=0.48\columnwidth]{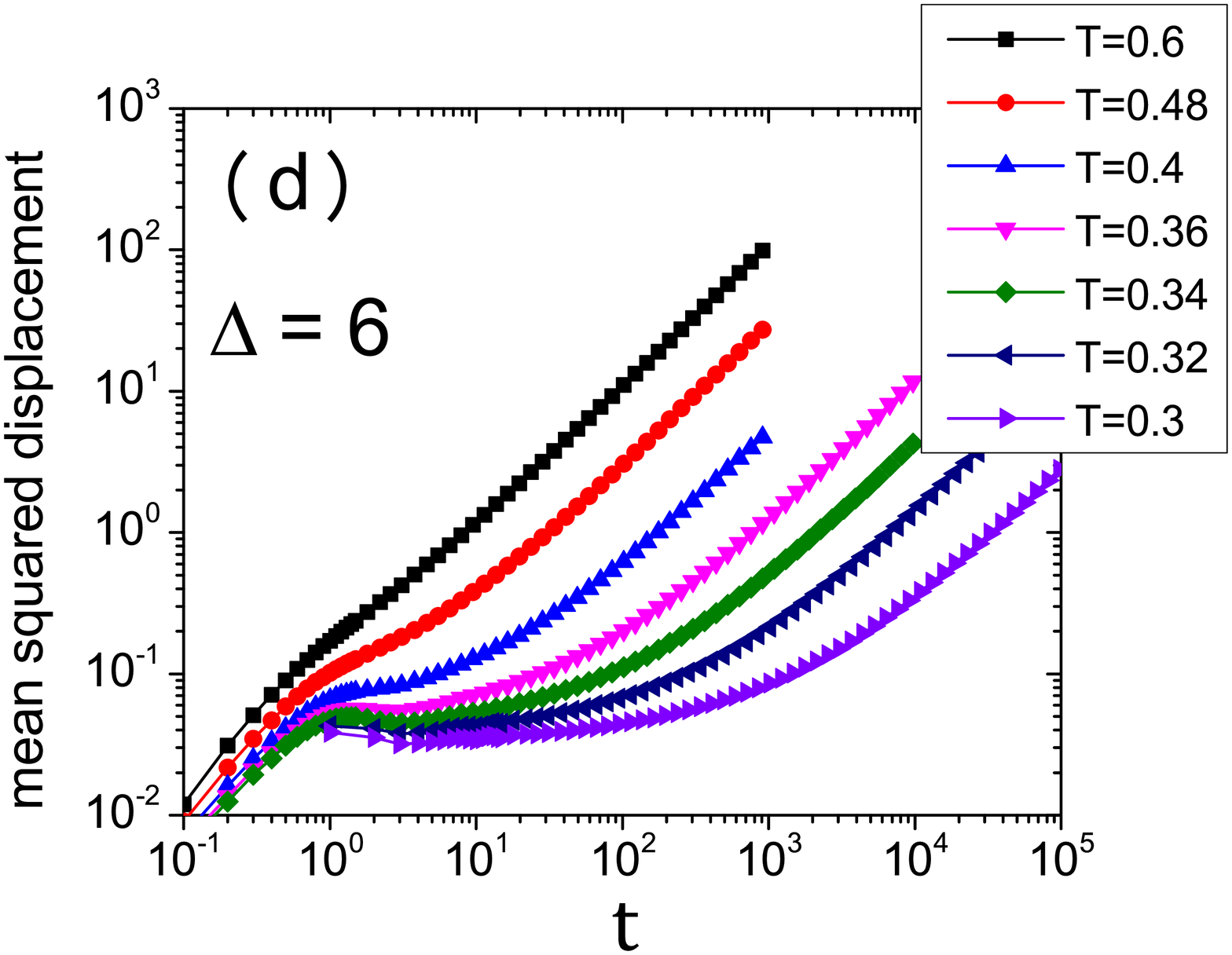}
\caption{
(a)--(d): The temperature variation of the mean squared displacement for species 1 for several $\Delta$'s.
}
\label{fig:MSD}
\end{center}
\end{figure}  

We also quantify the self diffusion constant $D_{\alpha}$ ($\alpha \in
\{ 1, 2 \}$) from the long-time behavior of the mean squared displacement (MSD).
The MSD for species 1 is presented in Figure~\ref{fig:MSD}.
Figure~\ref{fig:Arrhenius_plot} (b) shows $(D_1/T)^{-1}$ as a function of
$T_0/T$.

\begin{figure}
\begin{center}
\includegraphics[width=0.95\columnwidth]{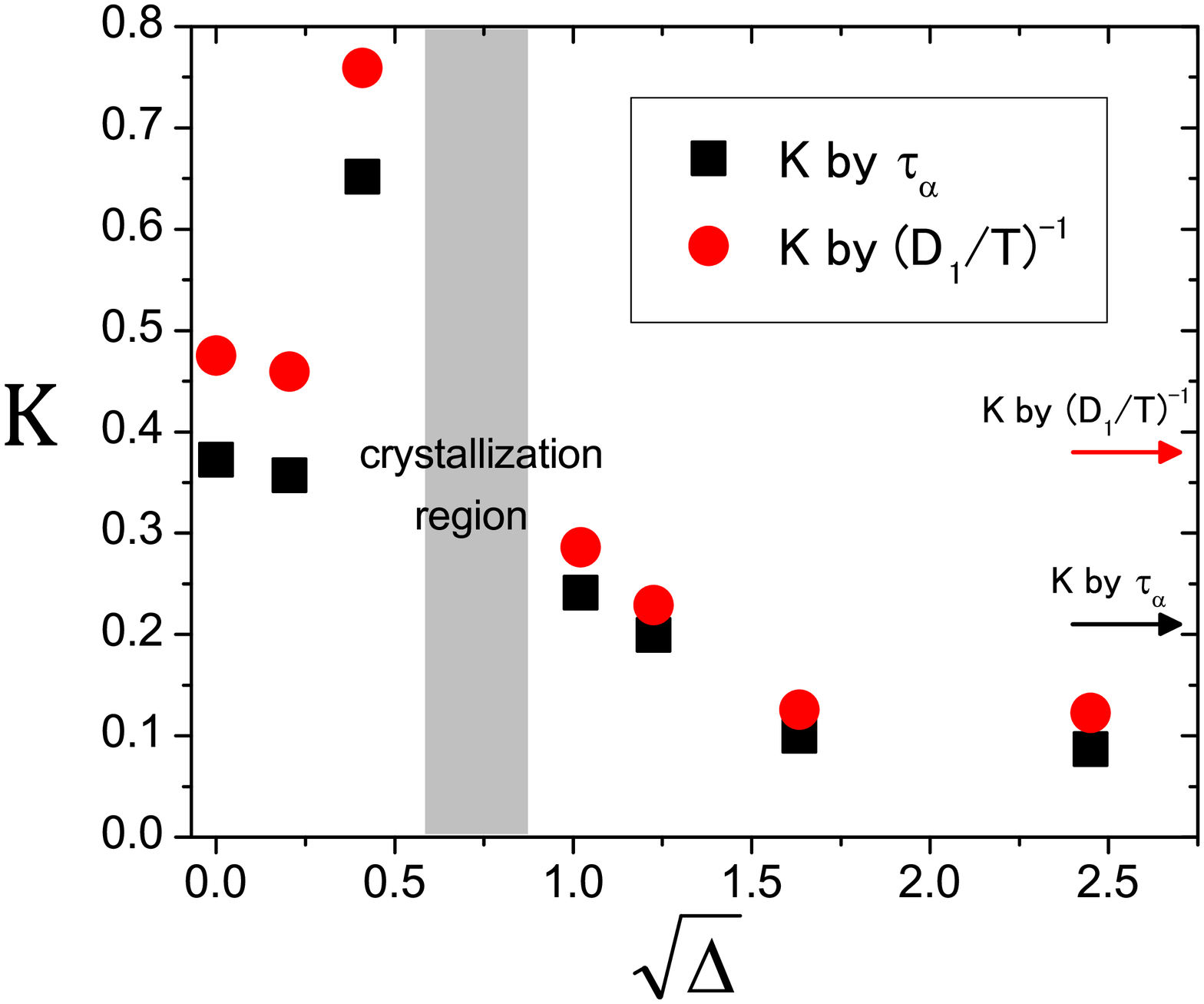}
\caption{
The fragility index $K$ as a function of $\Delta$.
The results of the VFT fitting by $\tau_{\alpha}$ and $(D_1/T)^{-1}$ are shown as the squares and
 circles, respectively.
The results of the Kob-Andersen mixture from Sengupta {\it et al.}\cite{sengupta2011dependence} are shown by the horizontal arrows.
The gray shaded area indicates the region where we observe crystallization at lower temperatures.
}
\label{fig:fragility}
\end{center}
\end{figure}

The $\Delta$ dependence of the fragility index $K$ is demonstrated 
in Figure~\ref{fig:fragility}.
At $\Delta=6$, the value of $K$ is small ($K \approx 0.1$), which is consistent
with the results of the previous study~\cite{coslovich2009dynamics}.
As shown in Figure~\ref{fig:fragility}, $K$ increases with decreasing
$\Delta$.
At $\Delta=1.042$ ($\sqrt{\Delta}=1.021$), $K$ exceeds the value of the Kob-Andersen (KA) mixture, which is 
a typical fragile glass former ($K\simeq 0.2$)~\cite{sengupta2011dependence}.
We address that the observed fragility index covers a wide range from $K= 0.087$ to $0.652$. 
Especially, the maximal value of $K=0.652$ at $\Delta=0.167$ is comparable with those of the most fragile
glass formers studied in simulations~\cite{malins2013identification,royall2015strong}.
$K$ is also obtained by using the
VFT equation for $(D_1/T)^{-1}$. 
The results are plotted with filled circles in Figure~\ref{fig:fragility}.
Although $K$ obtained from $(D_1/T)^{-1}$ are larger than
those obtained from
$\tau_{\alpha}$ for all $\Delta$'s, the overall behavior is qualitatively the same.
We remark that 
the increase of the fragility index with increasing the density has been reported for the BKS
model~\cite{van1990force}, although the investigated densities were limited~\cite{barrat1997strong}. 

The observation that the system becomes the most fragile ($K$ becomes the largest)
just next to the crystalline state may be related to the frustration
scenario which claims that the fragility is controlled by the frustration
against crystallization~\cite{tanaka2011roles}.
In fact, it has been reported that the system with a weaker frustration
against crystallization tends to have larger fragility indices~\cite{kawasaki2010structural,shintani2006frustration}.  
Further investigation of the dynamics around this Stishovite-like
crystallization regime would be valuable to verify the scenario.

\begin{figure}
\begin{center}
\includegraphics[width=0.95\columnwidth]{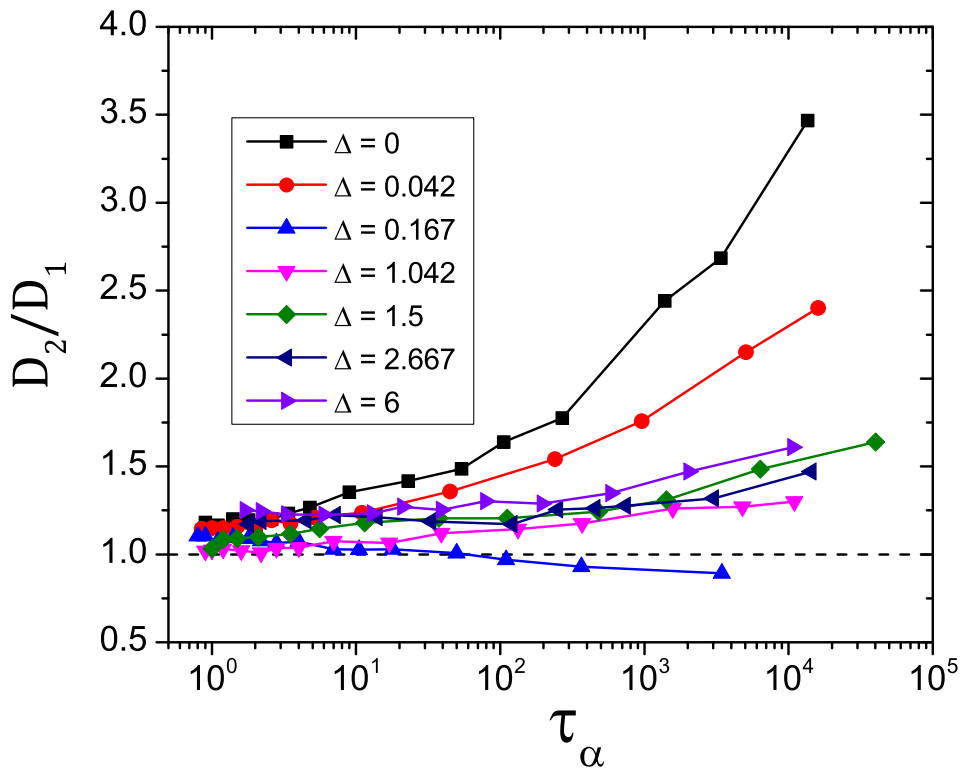}
\caption{
The ratio of the diffusion constants $D_2/D_1$ as a function of $\tau_{\alpha}$ for several $\Delta$'s.
}
\label{fig:decoupling}
\end{center}
\end{figure}

In order to characterize the difference of dynamics
between the strong (large $\Delta$) and fragile (small $\Delta$) regimes,
we evaluate the ratio of the diffusion constants $D_2/D_1$ between species 1 and 2.
In the experiments for $\rm SiO_2$, it has been reported that diffusion of silicon and oxygen atoms
decouples at low temperatures and the ratio of the diffusion constants for Si and O, 
$D_{\rm O}/D_{\rm Si}$, increases with 
decreasing the temperature~\cite{mikkelsen1984self}. 
Similar behavior has been also
demonstrated in simulations~\cite{horbach1999static,saksaengwijit2006origin}.
As mentioned in Reference~\cite{saksaengwijit2006origin},
this decoupling is attributed to the rotational motion of the oxygen atom in the tetrahedral structure.
In Figure~\ref{fig:decoupling}, we show $D_2/D_1$ as a function of
$\tau_{\alpha}$ for several $\Delta$'s.
At $\Delta=6$,  $D_2/D_1$ increases with increasing $\tau_{\alpha}$
(with decreasing the temperature).
This is consistent with the results for the original CP model~\cite{coslovich2009dynamics}.
As $\Delta$ decreases, variation of $D_2/D_1$ becomes milder and eventually becomes almost a flat line at $\Delta=0.167$. 
As $\Delta$ decreases further below $\Delta=0.167$,
the trend is reversed and $D_2/D_1$ becomes an increasing
function of $\tau_{\alpha}$ again and the slope keeps increasing until $\Delta$ reaches $0$, where the slope becomes maximum. 
The degree of the decoupling at $\Delta=0$, which corresponds to the SS model, is even larger than that of the network glass former 
at $\Delta=6$.
Similar strong decoupling between $D_1$ and $D_2$ 
has been reported for the conventional SS model~\cite{kawasaki2013slow}.
This non-monotonic $\Delta$ dependence of $D_2/D_1$
is another signature that the dynamics of the CP model
is changed qualitatively by tuning the potential depth $\Delta$.
It is natural to expect that the underlying mechanisms behind the strong
decoupling of $D_1$ and $D_2$ at the two extreme ends of $\Delta=0$ and $6$ are completely different. 
It would be worthwhile to seek for the origin of this strong decoupling
for $\Delta=0$.

\subsection{Absence of the density-temperature scaling}

The density-temperature (DT) scaling is an useful way to 
single out the parameter which controls the thermodynamic and dynamical properties of liquids.
The simplest example is the inverse power law (IPL) potential, $v(r) \sim r^{-n}$,
 where the DT scaling exactly holds and the system is characterized by a single
parameter, $\rho^{\frac{n}{d}}/T$, where $d$ is the spatial dimension~\cite{hiwatari1974molecular}.
It has been known that a broad class of liquids can be scaled by a
single parameter $\rho^{\gamma}/T$, where $\gamma$ is a constant 
~\cite{casalini2004thermodynamical,pedersen2008strong, schroder2009pressure3, gnan2009pressure4, schroder2011pressure5}.  
It has been argued that the DT scaling holds if there is a strong correlation between the virial $W$ and the potential energy $U$.
The liquids for which the correlation between $W$ and $U$ is more than $90\%$,
are called the {\it strongly correlating or Roskilde-simple liquids}~\cite{pedersen2008strong,
schroder2009pressure3, gnan2009pressure4, schroder2011pressure5}.
Since the CP model with $\Delta=0$ is the purely IPL system, the DT scaling exactly holds.
On the other hand, the previous study has demonstrated that at $\Delta=6$ (non-IPL system), the
correlation coefficient between $W$ and $U$ is less than $10\%$
(non-strongly correlating liquid), thus the DT scaling does not hold~\cite{coslovich2011heterogeneous}.
Consequently, the validity of the DT scaling for the CP model varies depending on $\Delta$. 

There is another type of the DT scaling which is
convenient to characterize dynamical quantities such as the
relaxation time $\tau_\alpha(\rho, T)$.
It has been argued that $\tau_\alpha(\rho, T)$ for many liquids is scaled by 
the characteristic time and energy at high temperatures, $\tau_\infty(\rho)$ and
$E_\infty(\rho)$.
$\tau_\infty(\rho)$ and $E_\infty(\rho)$ are determined by fitting $\tau_{\alpha}(\rho, T)$ at
high temperatures using the Arrhenius law;
\begin{equation}
\tau_{\alpha}(\rho, T) \simeq \tau_{\infty}(\rho)
 \exp[E_{\infty}(\rho)/T]. 
\label{eq:DT_scaling}
\end{equation}
This scaling is demonstrated to work well also in many supercooled liquids 
and polymer systems~\cite{alba2002temperature,
casalini2004thermodynamical, alba2006temperature}.

As mentioned in Section~\ref{sec:INTRODUCTION}, the density or pressure dependence of the fragility has been studied in various
systems~\cite{sastry2001relationship,de2004scaling,sengupta2013density,berthier2009compressing}. 
However, some of those results can be collapsed onto a master curve using the DT scaling, which means that the 
observed variation of the fragility is only superficial with no physical significance~\cite{de2004scaling, sengupta2013density}.

\begin{figure}
\begin{center}
\includegraphics[width=0.95\columnwidth]{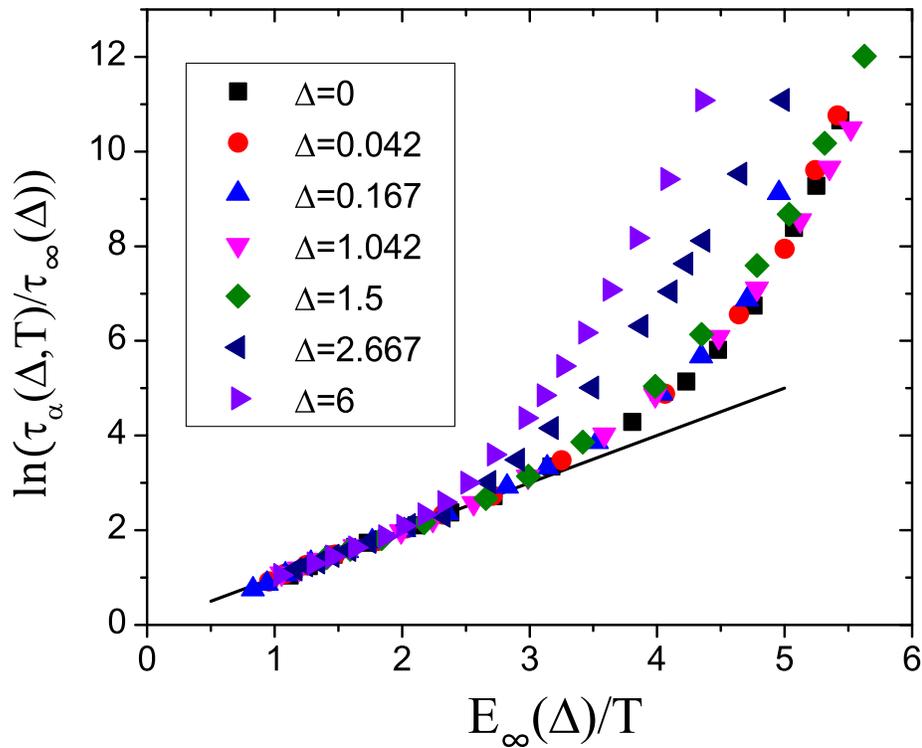}
\caption{
The density-temperature scaling for the relaxation time
 $\tau_\alpha(\Delta, T)$ scaled by $\tau_\infty(\Delta)$ as a function
 of $E_\infty(\Delta)/T$.
The straight line represents Equation~(\ref{eq:DT_scaling}).
}
\label{fig:DT_scalling_Tarjus}
\end{center}
\end{figure}

Here, we examine the DT scaling using Equation (\ref{eq:DT_scaling}), for the relaxation time
$\tau_\alpha(\Delta, T)$ of the CP model.
In Figure~\ref{fig:DT_scalling_Tarjus}, 
$\tau_\alpha(\Delta, T)$ scaled by $\tau_\infty(\Delta)$ is plotted as a
function of $E_{\infty}(\Delta)/T$.
In this figure, the potential depth $\Delta$ is used instead of the density $\rho$.
At the small $\Delta$ regime, the relaxation times are collapsed onto a master curve, whereas 
they systematically deviate from the master curve as $\Delta$ increases.
This result eloquently demonstrates that the dynamical properties of the CP model do not follow the DT scaling.
From these results, it is concluded that 
the observed variation of the fragility is not superficial but it is a genuine manifestation of the changeover of the physical mechanism.

\subsection{Stokes-Einstein violation and stretch exponent}

The advantage of the CP model is that the correlation of physical quantities 
with the fragility can be studied systematically in a single system over a wide range of fragility.
In particular, it enables one to examine the much debated issues on the relation of the fragility
with dynamical properties, such as the SE violation and the stretched exponential relaxation.
In normal liquids, it is expected that 
the SE relation, $D\eta/T \propto$ const. holds, where
$D$ and $\eta$ are the diffusion constant and viscosity, respectively.
However, the SE relation is violated in most supercooled liquids near the glass transition temperature.
This SE violation is often regarded as a manifestation of the spatially heterogeneous
dynamics, or {\it the dynamical heterogeneities}~\cite{ediger2000spatially}.
Previously, 
the SE violation of the original CP model ($\Delta=6$) has been investigated~\cite{kawasaki2014dynamics,staley2015reduced}.
These studies highlight the qualitative difference between the CP model 
($\Delta=6$) and other fragile glass formers
in their dynamical behavior~\cite{flenner2014universal}.
The aim of this section is to elucidate the dependence of the SE violation on the
fragility in a systematic way. 

\begin{figure}
\begin{center}
\includegraphics[width=0.48\columnwidth]{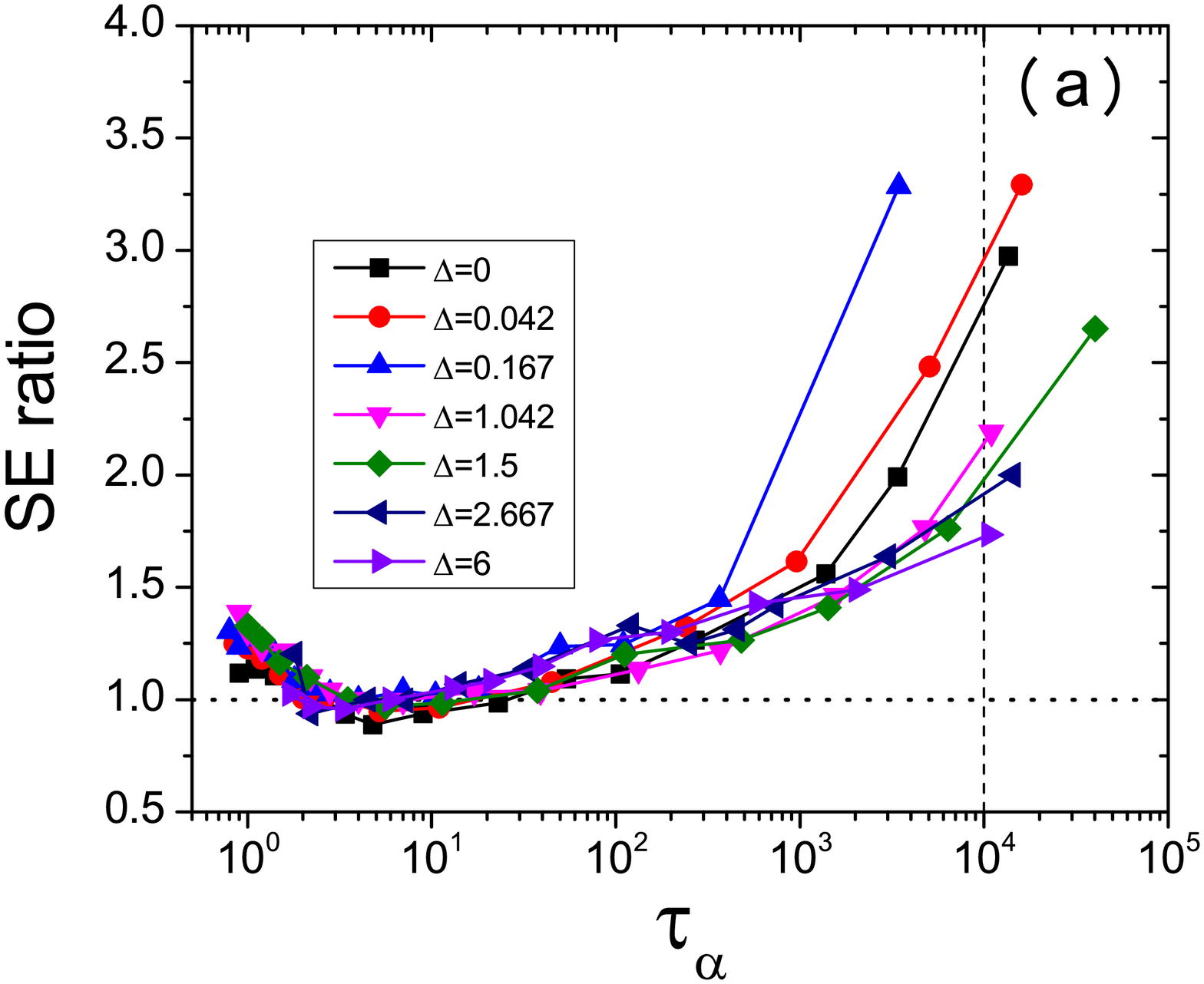}
\includegraphics[width=0.48\columnwidth]{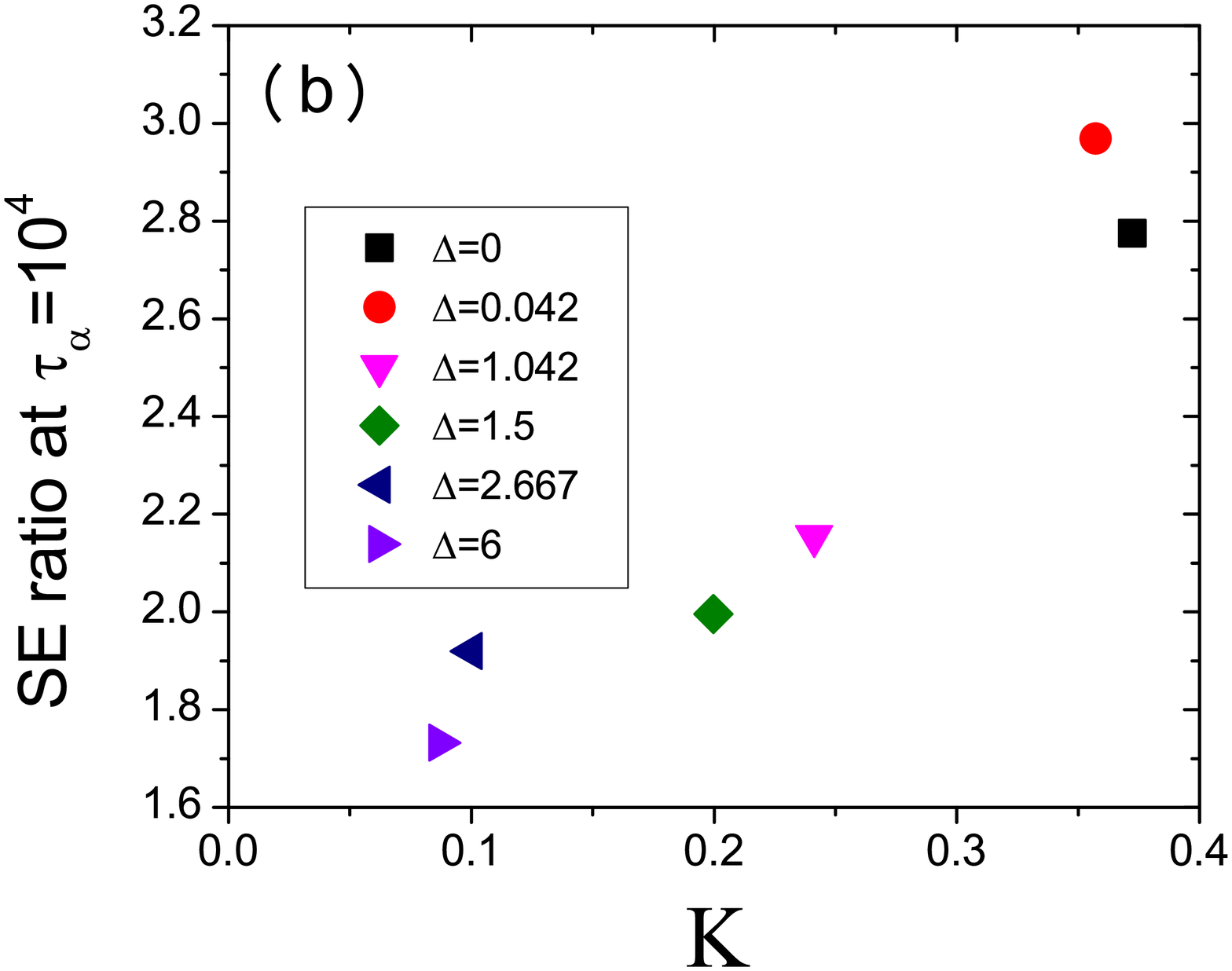}
\caption{
(a): The SE ratio as a function of $\tau_{\alpha}$.
The vertical dashed line corresponds to $\tau_{\alpha}=10^{4}$.
(b): The SE ratio at $\tau_{\alpha}=10^4$ as a function of $K$.
}
\label{fig:SE_violation}
\end{center}
\end{figure}

In Figure~\ref{fig:SE_violation}~(a), we show the Stokes-Einstein ratio, 
$D_1(T)\tau_{\alpha}(T)/D_1(T_{\rm onset}) \tau_{\alpha}(T_{\rm onset})$, normalized by the values at $T_{\rm
onset}$, as a function of $\tau_{\alpha}$.
For all $\Delta$'s, the SE relation is violated, \textit{i.e.}, the SE ratio increases with
increasing $\tau_{\alpha}$ (or decreasing the temperature).
We remark that the deviation of the SE ratio at high temperatures (small $\tau_\alpha$) is
an artifact caused by the use of $\tau_\alpha$ instead of
$\eta/T$~\cite{sengupta2013breakdown,shi2013relaxation,kawasaki2014dynamics}.
Dependence of the SE violation on the fragility is shown in Figure~\ref{fig:SE_violation} (b).
Here we plot the SE ratio at low temperatures (at $\tau_{\alpha}=10^4$) against the fragility
index $K$.
A clear correlation between the SE violation and the fragility
is observed, that is, the more fragile systems tend to exhibit the stronger SE violation.

Next, we study the stretch exponent $\beta_{\rm KWW}$ of the nonexponential relaxation observed
in the density
correlation function.
Experimental studies have shown that the fragile systems tend to have smaller stretch exponents than the strong systems~\cite{bohmer1993nonexponential,niss2007correlation}.
Some theories explain this
observation~\cite{xia2001microscopic}, while others argue that there is no direct correlation between the fragility and $\beta_{\rm KWW}$~\cite{dyre2007ten,heuer2008exploring}.

\begin{figure}
\begin{center}
\includegraphics[width=0.48\columnwidth]{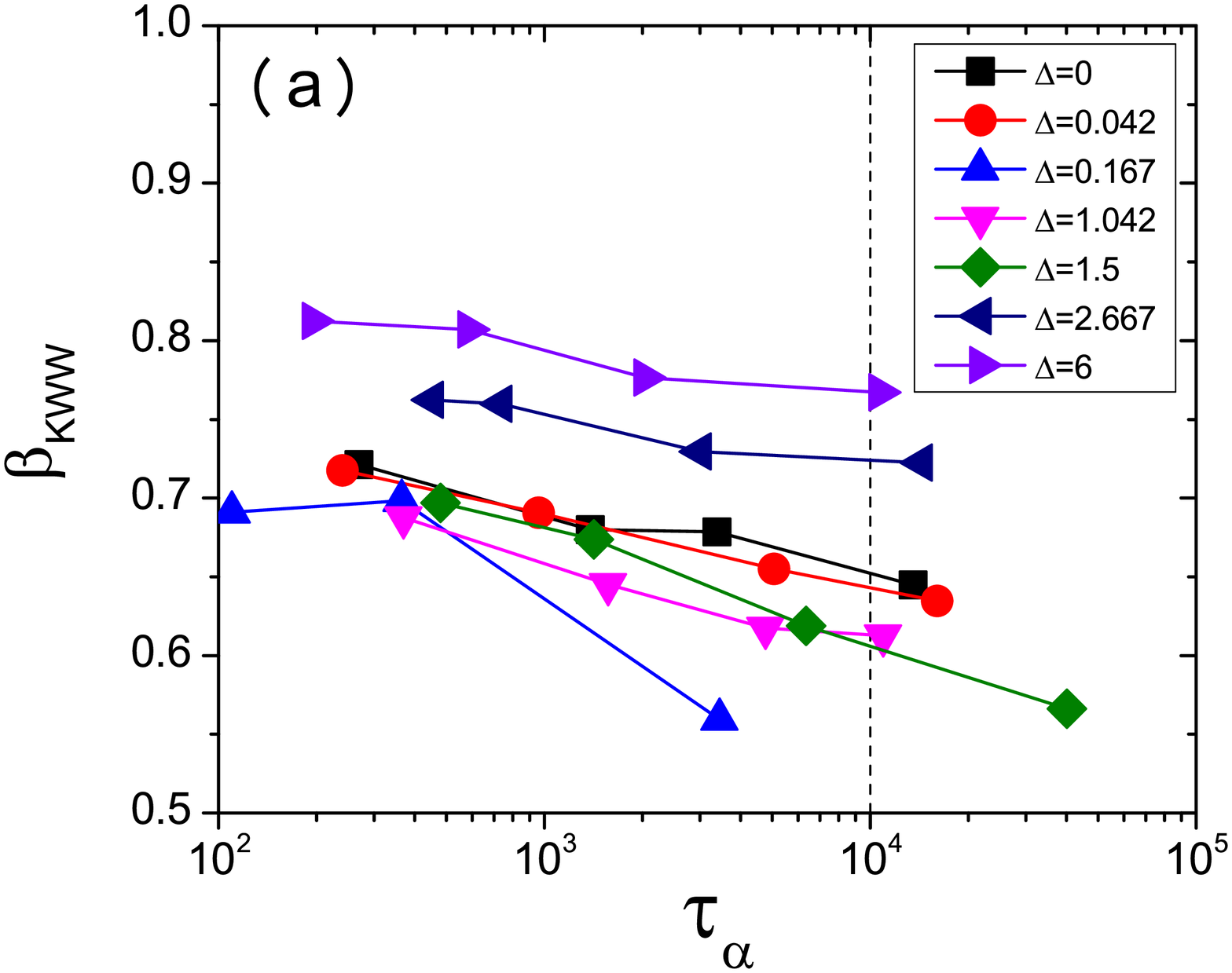}
\includegraphics[width=0.48\columnwidth]{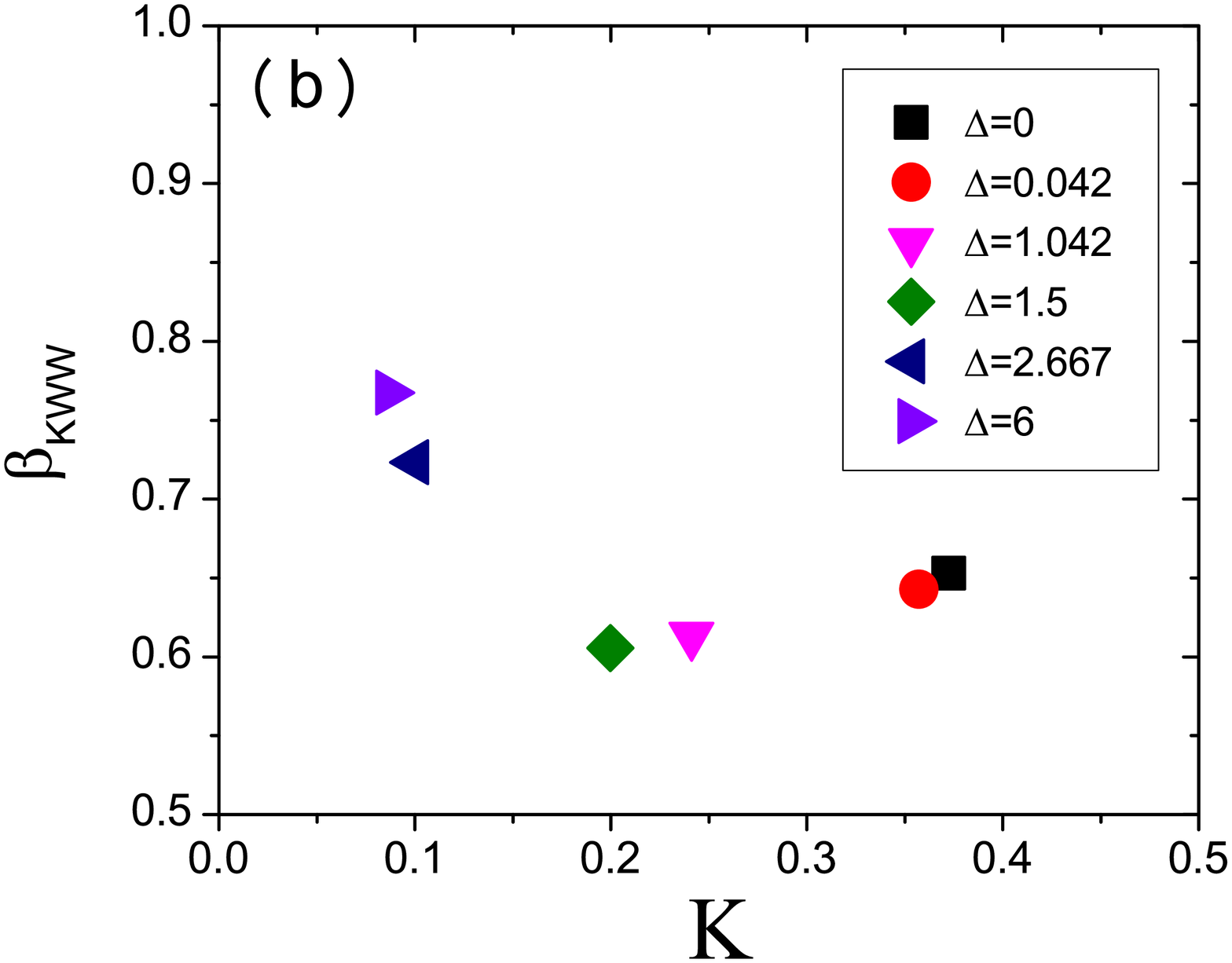}
\caption{
(a): The stretch exponent $\beta_{\rm KWW}$ as a function of $\tau_{\alpha}$.
The vertical dashed line corresponds to $\tau_{\alpha}=10^{4}$.
(b): $\beta_{\rm KWW}$ at $\tau_{\alpha}=10^{4}$ as a function of $K$.
}
\label{fig:stretched_exponent}
\end{center}
\end{figure}

Here, we show the fragility dependence of the stretch exponents of the CP model.
We determine $\beta_{\rm KWW}$ from $F_s(k, t)$ using the following fitting function,
\begin{equation}
F_s(k, t) = (1-f_{\rm c}) \exp[-(t/\tau_{\rm s})^2] + f_{\rm c} \exp[-(t/\tau_{\rm l})^{\beta_{\rm KWW}}],
\end{equation}
where $f_{\rm c}$, $\tau_{\rm s}$ and $\tau_{\rm l}$ are fitting parameters~\cite{sengupta2013breakdown}.
Figure~\ref{fig:stretched_exponent} (a) shows $\beta_{\rm KWW}$ as a function of $\tau_{\alpha}$.
For all $\Delta$'s, $\beta_{\rm KWW}$ decreases gradually with
increasing $\tau_{\alpha}$ (decreasing the temperature).
In Figure~\ref{fig:stretched_exponent} (b), we plot $\beta_{\rm KWW}$ at
$\tau_{\alpha}=10^4$ as a function of $K$.
Surprisingly, we observe no clear correlation between the fragility and $\beta_{\rm KWW}$, although
$\beta_{\rm KWW}$ becomes somewhat larger at strong-liquid regime of $\Delta=2.667$ and $6$. 
This observation is not inconsistent with the argument in Reference~\cite{heuer2008exploring}.
A similar trend has been reported for the harmonic
sphere model~\cite{berthier2009compressing}.

\subsection{Relationship with specific heat peak}

Finally, we discuss the possible relationship between the thermodynamic
properties and the fragility variation observed in the CP model.
Several simulation studies using the BKS model~\cite{van1990force} for
$\rm SiO_2$ have demonstrated that there exists the so-called fragile-to-strong (FS) crossover; 
the temperature dependence of the relaxation time changes from the
super-Arrhenius (fragile) to the Arrhenius (strong) behavior with
decreasing the temperature~\cite{horbach1999static,saika2001fragile,saika2004free}.
It has been argued that
this FS crossover is related with a thermodynamic anomaly signaled by appearance of the peak of the
specific heat~\cite{saika2001fragile,saika2004free,speck2014liquid} which 
tends to shift to lower temperatures if the density is increased~\cite{saika2001fragile,saika2004free,saika2004phase}.
This anomaly might be connected to a hidden thermodynamic singularity such as the liquid-liquid transition. But the overall picture is still unclear.
Under these circumstances, it is of great interest to examine whether the
FS crossover and the shift of the specific heat are also observed in the CP model
by tuning the potential depth $\Delta$.

\begin{figure}
\begin{center}
\includegraphics[width=0.95\columnwidth]{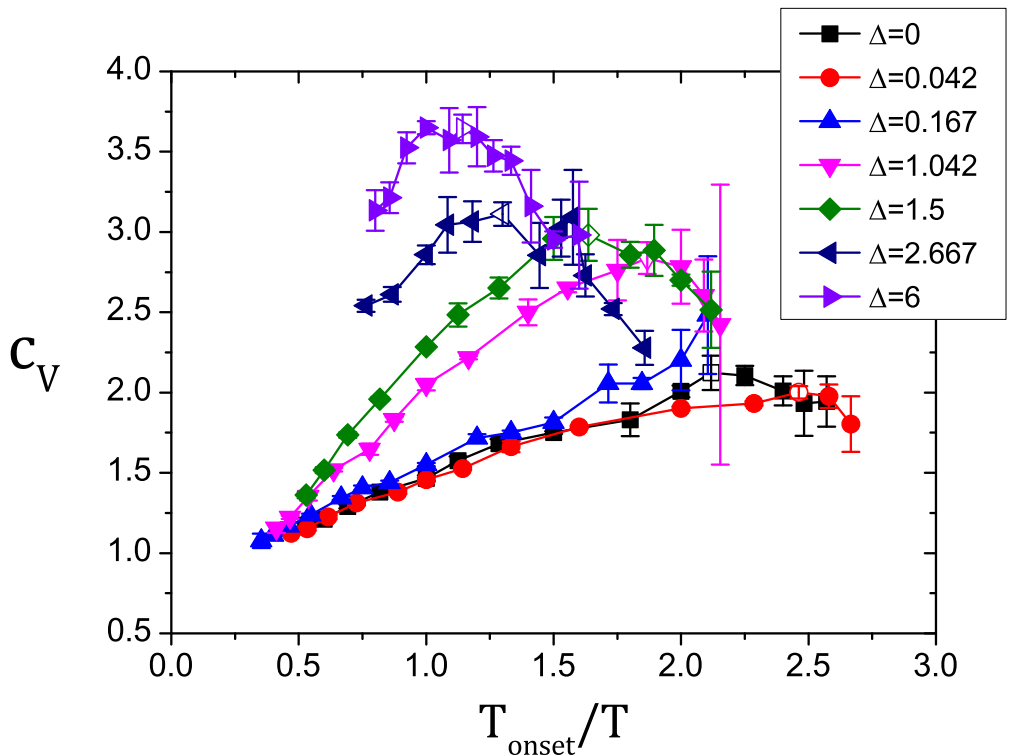}
\caption{
The specific heat as a function of the inverse temperature for several $\Delta$'s.
The temperature is scaled by $T_{\rm onset}$.
The open symbols are the positions of the peak.
}
\label{fig:specific_heat}
\end{center}
\end{figure}

The specific heat is calculated by $c_V=\frac{1}{N} \frac{\partial U}{\partial T}$.
Figure~\ref{fig:specific_heat} shows the temperature dependence of $c_V$
for various potential depth $\Delta$.
The temperature $T$ is normalized by the onset temperature $T_{\rm onset}$.
As observed,
there exists the broad but clear peak of $c_V$ at a finite temperature $T^*$ for all $\Delta$'s
except for $\Delta=0.167$.
We cannot access the lower temperatures for $\Delta =0.167$ because the system crystallizes.
The temperature $T^*$ for each $\Delta$ is listed in Table~\ref{tab:parameters}.
We find that $T^*$ shifts to lower temperatures with decreasing $\Delta$
(increasing the density).

Contrary to the distinct peaks of the specific heat, the FS crossover is harder to detect in the simulation data.
Due to the limited range of the temperature in the Arrhenius plot of Figure \ref{fig:Arrhenius_plot}, 
one hardly observe any sort of distinct crossover.
In this figure, we marked the temperatures at which the peak of the specific
heat are observed with the empty symbols for the guide of eyes.
For the data of $\Delta =2.667$ and 6, it is not impossible
to fit the data of $\tau_\alpha$ with the Arrhenius and super-Arrhenius laws on the low and high
temperature sides across $T^{\ast}$, but the fitting range is too narrow to call it convincing.
Similar fitting was also possible for smaller values of $\Delta$'s, but it is less trustworthy. 
Therefore, it would be fair to say that the existence of the correlation between the FS
crossover and the peak of the specific heat is not conclusive in current numerical simulations~\cite{saika2001fragile,saika2004free,saika2004phase}.
More extensive simulations are required to clarify the relation between the FS crossover and the specific heat peak.

\section{SUMMARY AND DISCUSSION}
\label{sec:SUMMARY_AND_DISCUSSION}

We have numerically investigated the CP model by tuning the depth
of the potential $\Delta$.
Changing $\Delta$ corresponds to changing the density of the system.
This approach has enabled us to perform
simulations over a wide range of density from the order of unity up to infinity. 
From the radial distribution functions, the coordination numbers, and
the static structure factors, we found
that the anisotropic tetrahedral network structures
are broken and then the isotropic structures are formed by decreasing $\Delta$ (or increasing the density).
We have also calculated various dynamical quantities such as the
relaxation time and diffusion constant.
These calculations have revealed
that their temperature dependence seamlessly changes from the Arrhenius to
the super-Arrhenius behavior.
We also confirmed that the temperature dependence of the dynamics 
is not collapsed by the DT scaling, assuring that the observed fragility change is
not a consequence of the trivial density scaling but due to the generic change of the mechanism of
the glassy dynamics.
We have studied the relationship between the
fragility and two dynamical quantities, the magnitude of the SE
violation and the stretch exponent of the density correlation
function.
The magnitude of the SE violation, which is believed to be a
manifestation of the dynamical heterogeneities, correlates well with the
fragility of the CP model.
On the contrary, the clear correlation between the fragility and the
stretch exponent has not been observed.
Finally, the peak of the specific heat and the possible correlation with the fragile-to-strong
(FS) crossover have been argued.
The peak has been observed in the temperature dependence of the
specific heat for most $\Delta$'s in the CP model.
In addition, its peak position shifts to lower temperatures with
decreasing the potential depth $\Delta$ (or increasing the density).
These observations are related with the recent numerical results
obtained in the BKS model~\cite{saika2001fragile,saika2004free,saika2004phase}.
However, it was difficult to identify the FS crossover temperature from the relaxation times
because the simulation time windows are too narrow and we conclude that the correlation between the FS crossover and
thermodynamic anomaly is still elusive. 
Further numerical investigation is required.

We address that the CP model exhibits a very broad variation of the
fragility, the highest and lowest values of which are comparable
to those of the most fragile and strongest liquids studied numerically in the past, and that it can serve as an ideal bench for numerical study of the fragility. 
Note that the harmonic sphere model also shows a very wide
variation of the fragility as the density is varied~\cite{berthier2009compressing,berthier2011role}.
This model also does not satisfy the DT scaling.
However, it should be emphasized that this model is qualitatively different 
from the CP model.
In the harmonic sphere model, the potential is truncated at a certain cutoff distance.
At low densities, the model becomes effectively a hard sphere system. 
Therefore, the temperature and energy are not relevant
but, instead, the pressure or density become the controlling thermodynamic parameters. 
This means that at low densities the temperature dependence of the relaxation time 
is trivially Arrhenius, whereas at higher densities, the temperature plays a pivotal role, leading to more
diverging (or fragile) behavior.
For the CP model, on the contrary, the temperature plays a role as a
relevant control parameter for all densities and the fragility is continuously controlled.

There are other
physical quantities that are expected to be related with the fragility~\cite{greer2014fragility};
the excess entropy~\cite{martinez2001thermodynamic}, the elastic
constants~\cite{novikov2004poisson, novikov2005correlation},
the non-ergodicity parameter~\cite{scopigno2003fragility}, the boson peak
frequency~\cite{novikov2005correlation}, and the temperature dependence of
microscopic structure~\cite{mauro2014structural}.
In addition, the relationship between the fragility and the dynamical
heterogeneities is a topic which attracts much attention recently~\cite{qiu2003length,berthier2005direct,kim2013multiple}.
The comprehensive information on the correlation of these properties with the fragility will be
obtained from the systematic study of the CP model by tuning a single parameter $\Delta$.
Further investigations along this line are currently underway.

\section*{Acknowledgments}
We thank D. Coslovich, A. Ikeda, H. Ikeda, and T. Kawasaki for helpful discussions.
M. O. acknowledges the financial support by Grant-in-Aid for Japan
 Society for the Promotion of Science Fellows (26.1878).
This work is partially supported by KAKENHI Grants No. 24340098 (K.M.),
 25103005 ``Fluctuation \& Structure'' (K.M.), 25000002 (K.M.), 26400428 (K.K.),
16H00829 ``Soft Molecular Systems'' (K.K.), and 26400428 (K.K.).
K. M. also thanks the JSPS Core-to-Core program.
The numerical calculations have been performed at Research Center for
Computational Science, Okazaki, Japan.

\section*{References}

\providecommand{\newblock}{}

\end{document}